\documentclass[prl,aps,reprint,footnotebib,amssymb,superscriptaddress]{revtex4-2}
\usepackage{amsmath}
\usepackage{bm}
\usepackage[dvipsnames]{xcolor}
\usepackage{graphicx}
\usepackage{epstopdf}
\usepackage{epsfig}
\usepackage{amsfonts}
\usepackage[pdfduplex=DuplexFlipLongEdge,colorlinks=true, citecolor=Blue, urlcolor=Blue, linkcolor=Blue, breaklinks=true, pdfencoding=auto,naturalnames]{hyperref}
\usepackage{mathtools}
\usepackage{verbatim}
\usepackage{tabularx}
\usepackage{bbm}
\usepackage{subfigure}
\usepackage{slashed}
\usepackage{physics}
\usepackage[normalem]{ulem}
\usepackage{array}
\usepackage{multirow}
\usepackage{times}
\newcommand{\myeqref}[1]{(\ref*{#1})} %

\def\bea{\begin{eqnarray}}
	\def\eea{\end{eqnarray}}
\def\be{\begin{equation}}
	\def\ee{\end{equation}}
\def\nn{\nonumber}
\def\ba{\begin{array}}
	\def\ea{\end{array}}
\def\Tr{\mathrm{Tr}}

\def\llangle{\langle \hspace{-0.23em} \langle}
\def\rrangle{\rangle \hspace{-0.22em} \rangle}
\newcommand{\ssll}{\hspace{-1em}---}

\begin{document}
	\title{Boundary criticality in two-dimensional interacting topological insulators}

	\author{Yang Ge}
	\affiliation{Department of Physics and Engineering Physics, Tulane University, New Orleans, Louisiana 70118, USA}
	
	\author{Hong Yao}
	\email{yaohong@tsinghua.edu.cn}
	\affiliation{Institute for Advanced Study, Tsinghua University, Beijing 100084, China}
	
	\author{Shao-Kai Jian}
	\email{sjian@tulane.edu}
	\affiliation{Department of Physics and Engineering Physics, Tulane University, New Orleans, Louisiana 70118, USA}
	
	\date{\today}
	
	\begin{abstract}
		We study the boundary criticality in 2D interacting topological insulators. 
		Using the determinant quantum Monte Carlo method, we present a nonperturbative study of the boundary quantum phase diagram in the Kane-Mele-Hubbard-Rashba model. 
		Our results reveal rich boundary critical phenomena at the quantum phase transition between a topological insulator and an antiferromagnetic insulator, encompassing ordinary, special, and extraordinary transitions. 
		Combining analytical derivation of the boundary theory with unbiased numerically exact quantum Monte Carlo simulations, we demonstrate that the presence of topological edge states enriches the ordinary transition that renders a continuous boundary scaling dimension and, more intriguingly, leads to a special transition of the Berezinskii-Kosterlitz-Thouless type. 
		Our work establishes a framework for the nonperturbative study of boundary criticality in two-dimensional topological systems with strong electron correlations.
		
	\end{abstract}
	\maketitle
	
	\paragraph{Introduction}\ssll
	Boundary physics plays a fundamental role in condensed matter physics, particularly in the study of topological phases of matter~\cite{castro2009the,hasan2010colloquium,qi2011topological,armitage2018weyl,lv2021experimental}. 
	In systems with nontrivial topology, the bulk-boundary correspondence guarantees the emergence of exotic boundary states that are crucial for understanding various quantum phenomena. 
	These include quantum Hall states~\cite{klitzing1980new,laughlin1981quantized,tsui1982two}, topological insulators~\cite{kane2005quantum,bernevig2006quantum,konig2007quantum}, and topological superconductors~\cite{schnyder2008classification,kitaev2009periodic,sato2017topological}, where gapless edge or surface states emerge due to the topological nature of the bulk. 
	A celebrated example is the helical edge state in a quantum spin Hall insulator, which is protected by time-reversal symmetry and remains robust against weak perturbations~\cite{kane2005Z2,wu2006helical,CKXu2006PRB,hsu2021helical}.

	Boundaries also play a significant role in critical phenomena. 
	The interplay between bulk phase transitions and boundary effects gives rise to a rich variety of boundary critical behaviors, a subject that falls under the framework of boundary conformal field theory (BCFT)~\cite{cardy1984conformal,burkhardt1994ordinary,liendo2013bootstrap,andrei2018boundary}. 
	Depending on the nature of the boundary conditions and interactions, different types of boundary criticality can emerge, including the ordinary transition, the special transition, and the extraordinary transition~\cite{domb1986phase,diehl1996the,cardy1996scaling}.
	For the ordinary transition, the boundary orders following the bulk upon the bulk quantum phase transition (QPT), whereas for the extraordinary transition, the boundary orders ahead of the bulk transition. 
	The point at which these two transitions meet defines the multicritical special transition.
	These classifications reflect the diverse ways in which boundary fluctuations interact with bulk criticality, leading to distinct scaling behaviors and universality classes.
	In BCFT, the primary field of the order parameter features distinct boundary scaling dimensions for different types of boundary criticality~\cite{domb1986phase}.
	
	Recently, the interplay between topology and criticality has garnered increasing attention~\cite{scaffidi2017gapless,zhang2017unconventional,wu2020boundary,verresen2021gapless,ma2022edge,yu2022conformal,yu2024universal,shen2024new}.
	For instance, it has been shown that the transition between the one-dimensional Haldane phase and a trivial phase that breaks the protecting symmetry is gapless on the boundary.
	In higher dimensions, theoretical studies have proposed that topological edge states can significantly influence the nature of boundary criticalities. 
	In particular, Ref.~\onlinecite{shen2024new} points out that in the transition from a topological insulator to a trivial insulator, the presence of nontrivial boundary states will modify the boundary scaling dimensions in an ordinary transition and, more intriguingly, give rise to a special Berezinskii-Kosterlitz-Thouless (BKT) transition. 
	However, these proposals have largely been based on perturbative approaches, and an unbiased, nonperturbative numerical confirmation remains lacking.
	
	In this work, we address this question by investigating a lattice model of a two-dimensional interacting topological insulator using large-scale determinant quantum Monte Carlo (DQMC) simulations~\cite{fehske2007computational,assaad2020ALF}. 
	Our model undergoes a bulk QPT from a topological insulator to an antiferromagnetic (AFM) state, belonging to the three-dimensional (3D) Ising universality class~\cite{zheng2011particle,li2017edge}. 
	Remarkably, we uncover a rich boundary phase diagram that includes both an ordinary and a special transition, which deviates from the boundary universality class of the 3D Ising model. 
	Specifically, our results demonstrate that the helical edge state enriches the ordinary transition, leading to a continuous boundary scaling dimension. 
	More strikingly, as the boundary interaction increases, the ordinary transition eventually culminates in a special BKT transition. 
	Our unbiased large-scale DQMC study establishes the boundary critical behavior in a two-dimensional interacting topological insulator.
	The boundary phenomena uncovered in our study pave the way for future explorations of boundary physics in topological and correlated quantum systems.

	\begin{figure}
		\includegraphics[width=0.9\linewidth]{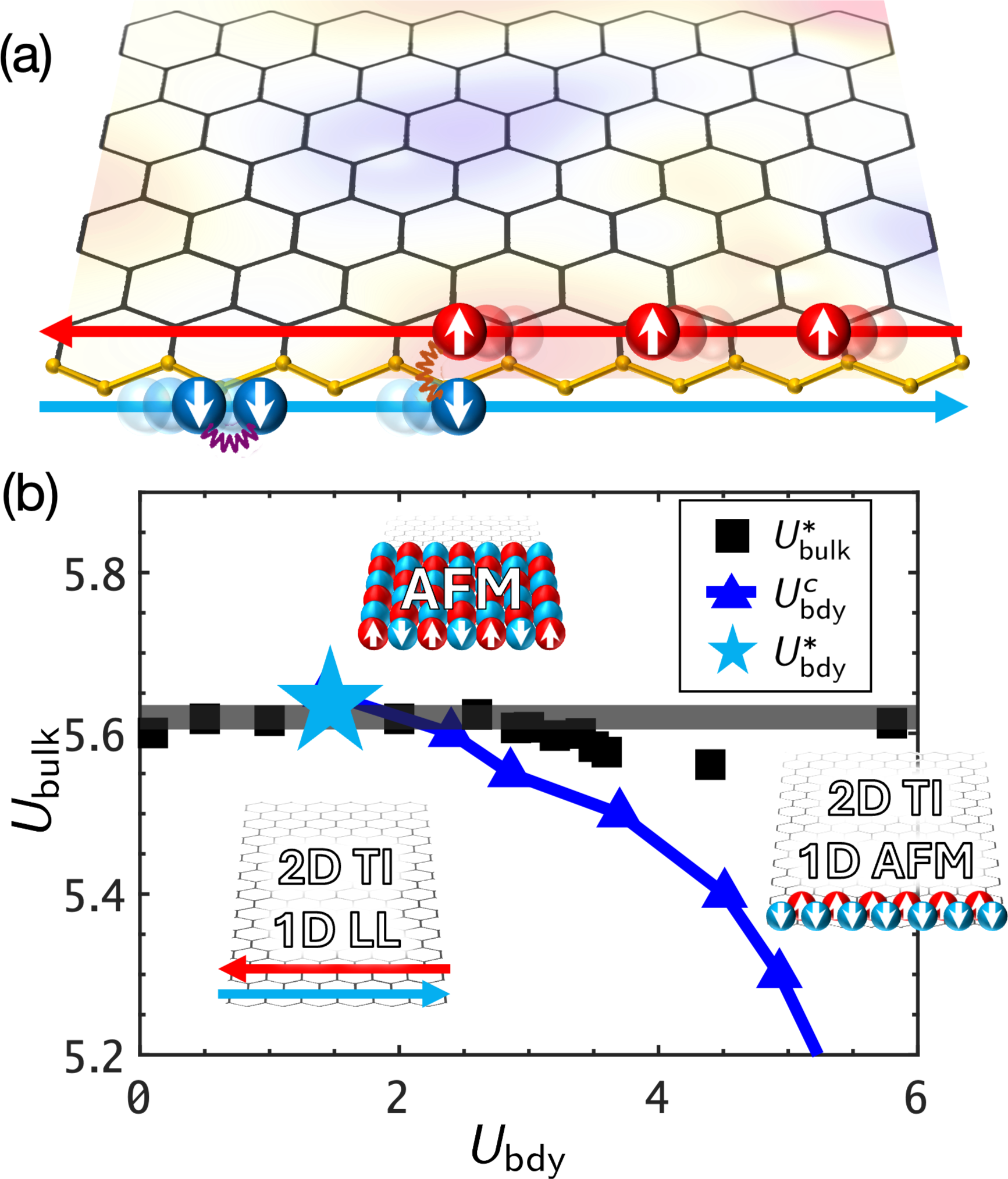}
		\caption{(a) Illustration of the lattice. Topological edge states reside on the zigzag boundary along with a tunable Hubbard interaction. (b) The quantum phase diagram showing bulk and boundary phases.
			The gray line shows the average $U^*_\text{bulk}$ for $U_\text{bdy}<2.5$. Scaling collapse is used to determine $U^*_\text{bulk}$, and $U^c_\text{bdy}$ at $U_\text{bulk}\le5.4$, while an extrapolation procedure is used for $U^c_\text{bdy}$ and $U^*_\text{bdy}$ at $U_\text{bulk} > 5.4$, as demonstrated in Fig.~\ref{fig:crossing}.}
		\label{fig:phase}
	\end{figure}
	
	\paragraph{Lattice model}\ssll
	We investigate a lattice model of a two-dimensional interacting topological insulator, placed on a cylinder that is periodic in the $x$-direction and open in the $y$-direction.
	The Hamiltonian comprises the noninteracting Kane-Mele model, which hosts a topological quantum spin Hall state, along with Hubbard \cite{zheng2011particle,hohenadler2011correlation,hohenadler2012qptkmh,hohenadler2013correlation,rachel2018review} and Rashba interactions~\cite{li2017edge}. 
	It is given as follows:
	\begin{eqnarray}
		\label{eq:hamiltonian}
		H &=& H_0 + H_1, \label{eq:H}\\%
		H_0 &=& -t \sum_{\langle \boldsymbol{ij} \rangle} c_{\boldsymbol{i},\sigma}^\dagger c_{\boldsymbol{j},\sigma} + \lambda \sum_{\llangle \boldsymbol{ij} \rrangle} (i  \nu_{\boldsymbol{ij}} c^\dagger_{\boldsymbol{i}}  \sigma^z  c_{\boldsymbol{j}} + \text{H.c.}),  \label{eq:H0}
	\end{eqnarray}
	\begin{eqnarray}
		\label{eq:Hint}H_1 &=& U_\text{bulk} \sum_{\boldsymbol{i} \in \text{bulk}} \left( c_{\boldsymbol{i},\uparrow}^\dagger c_{\boldsymbol{i},\uparrow}^{\phantom{\dagger}} - \frac12\right) \left( c_{\boldsymbol{i},\downarrow}^\dagger c_{\boldsymbol{i},\downarrow}^{\phantom{\dagger}} - \frac12\right) \nonumber \\ %
		&& + V \sum_{\langle \boldsymbol{ij} \rangle} \left( c_{\boldsymbol{i}, \uparrow}^\dagger c_{\boldsymbol{j}, \uparrow}^\dagger c_{\boldsymbol{i},\downarrow} c_{\boldsymbol{j},\downarrow} + \text{H.c.} \right) \nonumber \\ %
		&& + U_{\text{bdy}} \sum_{\boldsymbol{i} \in \text{bdy}} \left( c_{\boldsymbol{i},\uparrow}^\dagger c_{\boldsymbol{i},\uparrow}^{\phantom{\dagger}} - \frac12\right) \left( c_{\boldsymbol{i},\downarrow}^\dagger c_{\boldsymbol{i},\downarrow}^{\phantom{\dagger}} - \frac12\right). \label{eq:H1}
	\end{eqnarray}
	Here, \( c_{\boldsymbol{i},\sigma} \) ($c_{\boldsymbol{i},\sigma}^\dagger$) represents the fermion annihilation (creation) operator with spin \( \sigma \) at site \( \boldsymbol{i} \), while \( t \) denotes the nearest-neighbor hopping amplitude. 
	The Kane-Mele term with the amplitude \( \lambda \) induces the spin-orbit-coupled (SOC) hoppings, given by $c^\dagger_{\boldsymbol{i}}\sigma^z c_{\boldsymbol{j}} \equiv c^\dagger_{\boldsymbol{i},\sigma} [\sigma^z]_{\sigma\sigma'} c_{\boldsymbol{j},\sigma'}$, between the next-nearest neighbors, with a sign \( \nu_{\boldsymbol{ij}} = \pm \) when the fermion propagates to the next-nearest neighbor via a left or right turn, respectively. 
	The interaction terms are \( U_\text{bulk} \) the bulk Hubbard interaction, \( U_\text{bdy} \) the boundary Hubbard interaction, and \( V \) the nearest-neighbor Rashba interaction.
	Note that while the SOC reduces the global spin $\mathrm{SU}(2)$ to the $\mathrm{U}(1)$ associated with total $S^z$ conservation, the Rashba interaction $V$, as introduced in Ref.~\onlinecite{li2017edge}, further breaks the $\mathrm{U}(1)$ explicitly, down to the $\mathbb{Z}_2$ time-reversal symmetry.
	It enables two-particle spin-flip scattering at the boundary, which is essential to opening a gap in the helical edge state at the boundary criticality as we will elaborate in the boundary theory.
	The summation \( \sum_{\boldsymbol{i} \in \text{bulk}} \) runs over the sites in the bulk, while \( \sum_{\boldsymbol{i} \in \text{bdy}} \) is restricted to the sites on the boundary. The model is illustrated in Fig.~\ref{fig:phase}(a) with boundary sites highlighted.
	
	In the topologically nontrivial regime, the Kane-Mele model $H_0$ hosts helical edge states that are protected by time-reversal symmetry~\cite{kane2005Z2}. As the interaction strength $U_\text{bulk}$ increases, the system undergoes a QPT from a topological insulator to an AFM state.
	In the AFM phase, the time-reversal symmetry is broken, leading to a gap opening in the helical edge states. 
	However, at the quantum critical point (QCP), the time-reversal symmetry remains preserved. 
	To investigate the interplay between the topological edge states and critical fluctuations, we introduce a tunable boundary interaction $U_\text{bdy}$, which allows us to explore the impact of boundary correlations on the phase transition. 
	Importantly, the lattice model is free from the sign problem~\cite{li2015majorana,li2016majorana,li2017edge}, which would show up \cite{hohenadler2014rashba} in the presence of a noninteracting Rashba SOC \cite{kane2005quantum,bychkov1984oscillatory}. Thus, we are able to perform large-scale DQMC simulations to map the quantum phase diagram at zero temperature.
	The details of our numerical results are presented in the rest of the paper.

	\paragraph{Quantum phase diagram}\ssll
	The quantum phase diagram in Fig.~\ref{fig:phase}(b) is obtained through DQMC.
	We consider a lattice geometry with $L \equiv L_x = 2 L_y$, in which $L_{x,y}$ denotes the number of unit cells along the $x$- and $y$-axes, respectively. 
	The parameters are set to $t=1$, $\lambda = 0.3$ and $V=0.1$. 
	The simulated sizes range from $L=12$ to $L=24$ at inverse temperatures ranging from $\beta = 85$ to $\beta =120$
	\footnote{The ground state is degenerate and thus naively precludes the use of the projective DQMC. See the Supplemental~\cite{supplemental}.}.
	To explore the boundary criticality, we vary both $U_\text{bulk}$ and $U_\text{bdy}$ across the phase space. 
	Below, we discuss in detail the procedure for obtaining the phase diagram in Fig.~\ref{fig:phase}(b).
	
	As shown in Fig.~\ref{fig:phase}(b), the bulk transition from the TI to the AFM occurs at $U_\text{bulk} = U^\ast_\text{bulk}$.
	The ordering lies along $S^y$ due to the phase factor chosen in the Rashba term in Eq.~\eqref{eq:H1}.
	The critical point is determined by the crossing of the RG-invariant quantity of $S^y$ correlators, denoted by $R^{SS}$. It is defined as
	\begin{eqnarray}
		R^{SS}_\mathrm{w} = \frac1{2\pi} \sqrt{\frac{2 \tilde C^{SS}_\mathrm{w}(0)}{\tilde C^{SS}_\mathrm{w}(k_{\rm min}) + \tilde C^{SS}_\mathrm{w}(-k_{\rm min})} -1 }\,,  
	\end{eqnarray}
	where $\mathrm{w} \in \{\text{bulk}, \text{bdy}\}$ denotes the bulk or boundary quantities, respectively, and $k_{\min} = \frac{2\pi}L$. $\tilde C_\mathrm{w}^{SS}(k) $ is the Fourier transform of the AFM correlation function,
	\begin{eqnarray}
		\tilde C_\mathrm{w}^{SS}(k) &=& \frac1{\sqrt{L}} \sum_r C_\mathrm{w}^{SS}(r) e^{- i k r}, \\
		C_\mathrm{w}^{SS}(r) &=& \frac1{N_{\rm w}} \sum_{ \boldsymbol{i} \in \rm {w}} (-1)^{s(\boldsymbol{i}) + s(\boldsymbol{i}+\boldsymbol{r})} \langle S^y_{\boldsymbol i} S^y_{\boldsymbol{i} + \boldsymbol{r}} \rangle,
	\end{eqnarray}
	with $S^y_{\boldsymbol i} \equiv c^\dagger_{\boldsymbol i \sigma} [\sigma^y]_{\sigma\sigma'} c_{\boldsymbol i \sigma'}$, $\boldsymbol{r} = (r,0)$ being the displacement between two spin operators displaced along the $x$-axis, and $s(\boldsymbol{i}) = 0, 1$ for the two sublattices. 
	The normalization factors for the bulk and boundary quantities are given by $N_\text{bulk} = 2L^2 $ and $N_\text{bdy} = L $, respectively.
	
\begin{figure}
		\includegraphics[width=0.95\linewidth]{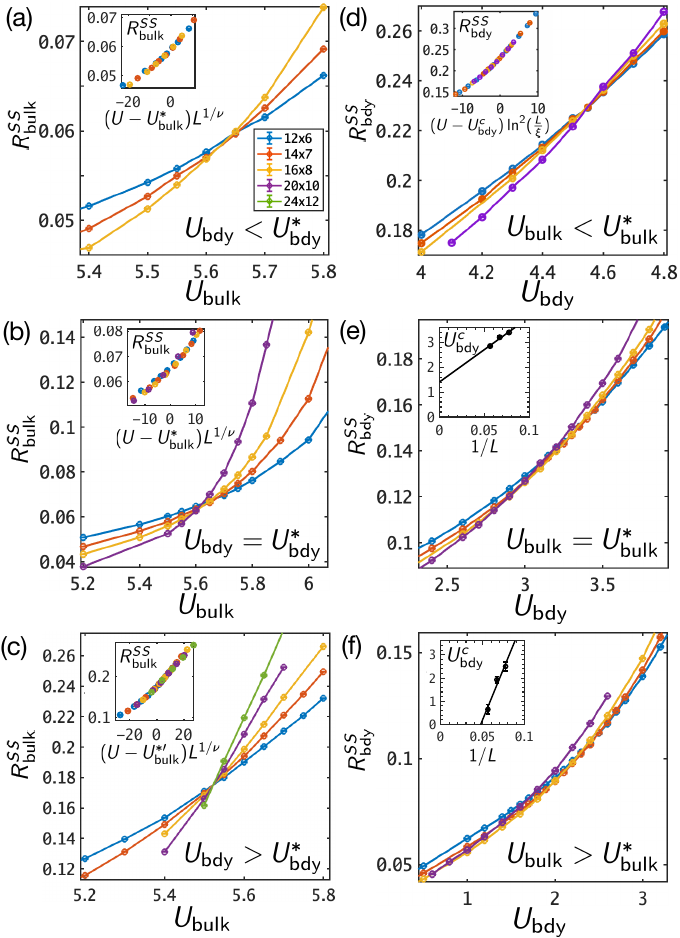}
		\caption{\label{fig:crossing}
			(a)--(c) The bulk RG-invariant quantity $R^{SS}_\text{bulk}$ showing crossings and finite-size scaling collapses (insets) at the bulk criticality, at (a) $U_\text{bdy}=0.5$, (b) $U_\text{bdy}=1.4$, and (c) $U_\text{bdy}=4.4$. For the collapses, $1/\nu=1.587$, and $U^*_\text{bulk}=5.65$ except for (c), where the crossing from the scaling analysis is lowered to $U^{\ast\prime}_\text{bulk}=5.53$ due to finite-size effects.
			(d)--(f) The boundary RG-invariant quantity $R^{SS}_\text{bdy}$ at (d) $U_\text{bulk}=5.4$, (e) $U_\text{bulk}=5.65$, and (f) $U_\text{bulk}=5.8$. The inset in (d) shows the collapse with logarithmic finite-size scaling of the driving field when the bulk is disordered, with $U^c_\text{bdy}=4.56$ and $\xi=0.42$. When the bulk is nearly critical or ordered, $R^{SS}_\text{bdy}$ has a strong finite-size effect, and the crossing at $U^c_\text{bdy}$ shifts significantly with $L$. Insets of (e) and (f) show linear extrapolations to $1/L \to 0$.}
	\end{figure}

	When $R^{SS}_\mathrm{w}$ is plotted as a function of the interaction strength, the data curves of different system sizes cross at the critical point, as shown in Fig.~\ref{fig:crossing}, which can drift due to finite-size effects.
	In our discussion, we denote the critical strength when the bulk is critical by $U^*_\mathrm{w}$, while that for the boundary transition when $U_\text{bulk}<U^*_\text{bulk}$ is denoted by $U^c_\text{bdy}$. They delineate the phase boundaries in Fig.~\ref{fig:phase}. 
	We use ${}^*$ to denote quantities at the bulk criticality, including those at the special transition.
	
	For the bulk transition, the data collapses in Fig.~\ref{fig:crossing}(a)--(c) confirm that the QPT belongs to the 3D Ising universality class with critical exponents $1/\nu = 1.587$.
	The collapse utilizes the finite-size scaling ansatz, $R_\text{bulk}^{SS} \sim f\left[ (U_\text{bulk} - U_\text{bulk}^\ast) L^{1/\nu}\right]$, for some smooth function $f$~\cite{autoscale,houdayer2004collapse}. 
	In the thermodynamic limit, the bulk transition point remains largely unaffected by $U_\text{bdy}$, as reflected in the consistency of the bulk critical points in Figs.~\ref{fig:crossing}(a) and (b). At larger $U_\text{bdy}$, finite-size effects become visible in Fig.~\ref{fig:crossing}(c): the crossing point shifts to lower $U_\text{bulk}$ due to the ordering at the boundary discussed next.
	
	For a fixed bulk interaction strength $U_\text{bulk}$, we vary the boundary interaction strength $U_\text{bdy}$ to determine the fate of helical edge states.
	It is probed by the boundary RG-invariant quantity $R^{SS}_\text{bdy}$.
	Before the bulk orders, i.e.\ at $U_\text{bulk} < U^\ast_\text{bulk}$, there is a boundary transition driven by $U_\text{bdy}$ indicated by the crossing in $R^{SS}_\text{bdy}$ for different system sizes as shown in Fig.~\ref{fig:crossing}(d). 
	This transition belongs to the BKT universality class of helical states where the Luttinger parameter $K = \frac12 $, as studied in Refs.~\onlinecite{zheng2011particle,li2017edge}. 
	The inset shows the data collapse of the RG-invariant quantity, consistent with the logarithmic scaling ansatz for the BKT transition, $R_\text{bdy}^{SS} \sim f\big[ (U_\text{bdy} - U^{c}_\text{bdy}) \log^2(L/\xi)\big]$. 
	Here, $U^{c}_\text{bdy}$ is the critical point of the boundary transition, generally dependent on the $U_\text{bulk}$, and $\xi$ is a fitting parameter in the data collapse. 
	On the other hand, after the bulk orders, i.e.\ at $U_\text{bulk} > U^\ast_\text{bulk}$, the boundary orders as well and the helical boundary states open a gap. Consequently, the crossing point shifts significantly and monotonically to smaller $U_\text{bdy}$ as the system size increases, leading to a negative $U_\text{bdy}$ in the extrapolated crossing points of $R_\text{bdy}^{SS}$, shown in the inset of Fig.~\ref{fig:crossing}(f). 
	
	Our main focus is the boundary criticality at the bulk QCP, $U_\text{bulk} = U_\text{bulk}^\ast$, where the bulk degrees of freedom are also gapless.  
	At this point, we observe that the $U_\text{bdy}$ of the $R_\text{bdy}^{SS}$ crossing point decreases monotonically with increasing system size.
	In the thermodynamic limit, it extrapolates to $U_\text{bdy}^\ast$ in Fig.~\ref{fig:crossing}(e). 
	This boundary critical point marks three distinct boundary critical behaviors: the ordinary transition for $U_\text{bdy} < U_\text{bdy}^\ast$, the extraordinary transition for $U_\text{bdy} > U_\text{bdy}^\ast$, and the special transition at $U_\text{bdy} = U_\text{bdy}^\ast$.
	
	\paragraph{Boundary theory}\ssll
	To understand the boundary criticality, we derive the helical edge state of the lattice model.
	We leave the details to the Supplemental Material~\cite{supplemental}, and present the Hamiltonian projected onto the helical edge state: 
	\begin{eqnarray}
		\hat H_\text{bdy} =   \frac{\tilde v}2 \left[ K \Pi^2 + \frac1{K} (\partial_x \varphi)^2 \right] + g_2 \cos (2 \sqrt{4\pi} \varphi), 
	\end{eqnarray}
	where $\varphi$ is the boson field originated from the bosonization of the helical edge state.
	The velocity and the Luttinger parameter are, respectively,
	\begin{equation}
		\label{eq:luttinger}
		\tilde v = v \sqrt{\left( 1 - \frac{U_\text{eff}}{2\pi v} \right) \left( 1 + \frac{U_\text{eff}}{2\pi v} \right)}, \ \   K = \sqrt{\frac{ 1 - \frac{U_\text{eff}}{2\pi v}}{ 1 + \frac{U_\text{eff}}{2\pi v}}}, 
	\end{equation}
	and $g_2 \propto V$ is the Rashba-interaction-induced two-particle scattering between the two spin components~\footnote{Our labels of $g_i$ couplings are different from the other bosonization conventions.}.
	The bare Fermi velocity and the effective interaction strength are derived in the Supplemental Material~\cite{supplemental} following~\cite{doh2013bifurcation}, and summarized below
	\begin{eqnarray}
		v &=&  \frac{6 \lambda t}{\sqrt{t^2 +(4\lambda)^2}}\,, \\
		U_\text{eff} &=& \frac{t^2(t^2 + 8 \lambda^2)(\sqrt{t^2 + (4\lambda)^2} - t)^4 }{ 4096 \lambda^8} (U_\text{bdy} - U_\text{bulk}) \nn \\
		&& + \frac{t}{\sqrt{t^2+(4\lambda)^2}} U_\text{bulk}\,.
	\end{eqnarray}
	The Luttinger parameter decreases as the boundary Hubbard interaction increases. 
	The two-particle scattering becomes relevant at $K = \frac12$, leading to a boundary transition when the bulk is disordered and preserves the time-reversal symmetry. 
	
	\begin{figure}
		\centering
		\includegraphics[width=\linewidth]{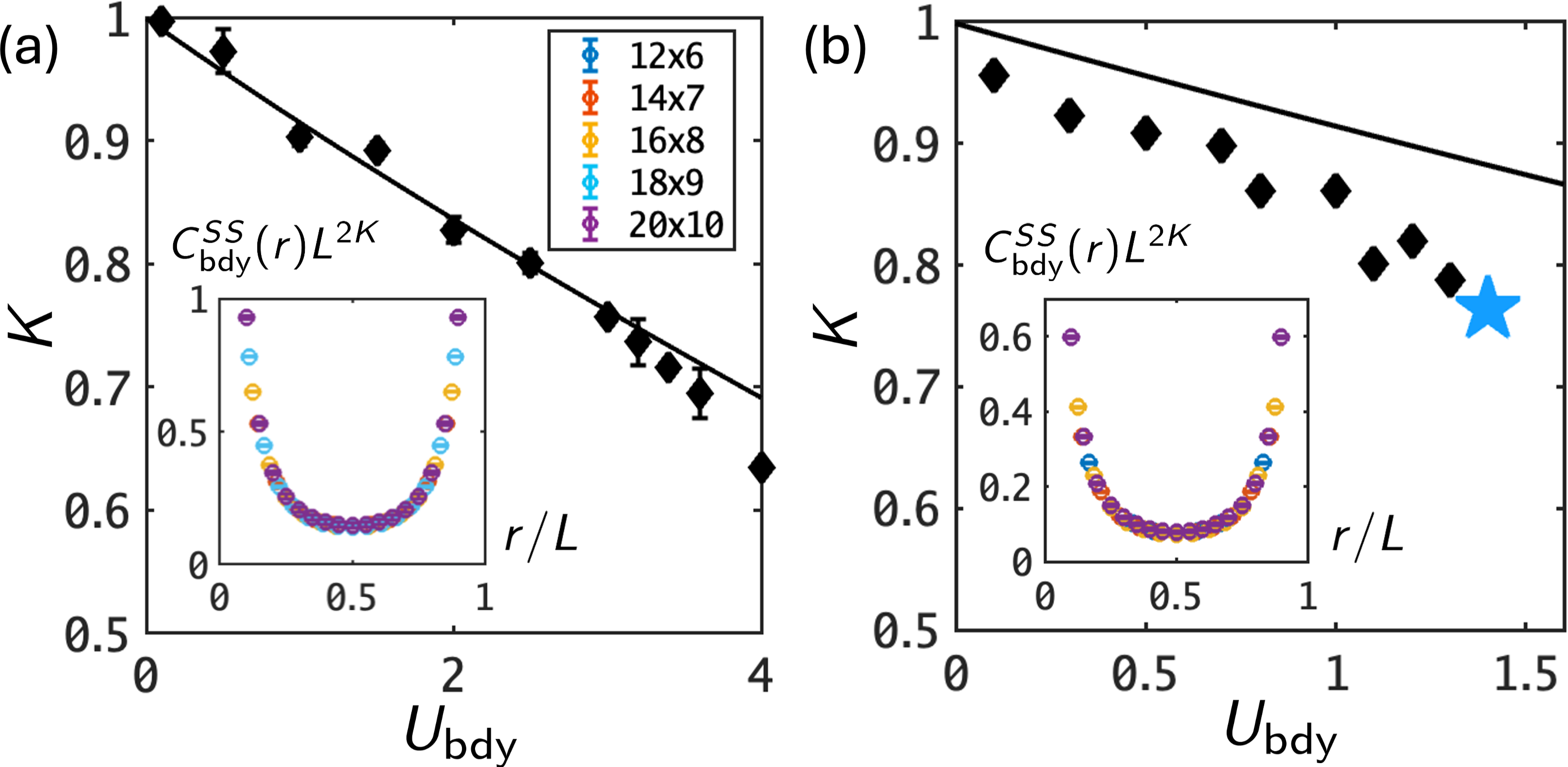}
		\caption{Luttinger parameter of the helical edge mode for (a) $U_\text{bulk} = 2.8 < U_\text{bulk}^\ast$, and (b) $U_\text{bulk} = 5.65 = U_\text{bulk}^\ast$.
			The black points are the extracted Luttinger parameter, while the solid line is the analytical prediction in \eqref{eq:luttinger}.
			The insets show the data collapse of the spin-spin correlation on the boundary at (a) $U_\text{bdy}=2$, and (b) $1$.
			The star denotes the special transition.}
		\label{fig:luttinger}
	\end{figure}
	This analytical result agrees well with the Luttinger parameter extracted from the DQMC calculations away from the bulk phase transition.
	Fig.~\ref{fig:luttinger}(a) shows the fitted Luttinger parameter at $U_\text{bulk} = 2.8 < U_\text{bulk}^\ast$ using the conformal ansatz
	$C_\text{bdy}^{SS}(r) L^{2K}  = f\left( \sin \frac{\pi r}{L} \right)$ for some smooth function $f$. 
	It is clear that the fitted Luttinger parameter agrees with the analytical prediction in~\eqref{eq:luttinger}. 
	
	\paragraph{Boundary criticality}\ssll
	At the transition $U_\text{bulk}^\ast$, the bulk of the system becomes gapless, and is described by a 3D Ising universality class. 
	The 3D Ising BCFT is well studied in nontopological systems.
	In particular, the leading boundary primary operator that carries the Ising symmetry is the order parameter field, denoted $\hat \phi$~\footnote{More precisely, at the ordinary transition, the boundary primary field is the normal derivative of the order parameter field. Here, we denote it as $\hat \phi$ for simplicity.}. 
	In the topological system, specifically the topological insulator considered here, we can imagine stripping off the outermost zigzag boundary sites where the interaction strength is $U_\text{bdy}$, along with the helical edge states.
	Without this outermost layer, the 3D Ising BCFT exhibits the ordinary transition since the interaction strength becomes uniform~\cite{deng2003bulk,deng2005surface}. 
	Then, imagine reattaching the zigzag boundary layer. This leads to a coupling between the helical edge state and the boundary order parameter field~\cite{shen2024new}:
	\begin{eqnarray}
		\hat H &=& \frac{\tilde v}2 \left[ K \Pi^2 + \frac1{K} (\partial_x \varphi)^2 \right] + g_2 \cos (2 \sqrt{4\pi} \varphi) \nn \\
		&& + g_1 \hat \phi \cos (\sqrt{4\pi} \varphi) \, ,
	\end{eqnarray}
	in which $\hat \phi$ denotes the boundary order parameter field that couples to the spin operator $S_y \propto \cos \left( \sqrt{4\pi} \varphi\right)$ in the helical Luttinger liquid. 
	We emphasize that while $\hat \phi$ would correspond to the boundary primary field in the 3D Ising universality class, in our present theory it serves as the bulk-boundary coupling field.
	The true boundary primary field in our case is the boundary spin operator $S_y$.

	As we argued above, $\hat \phi$ belongs to the ordinary universality class, with a scaling dimension $\Delta_{\hat \phi}$. 
	This scaling dimension has been obtained via various methods, including Monte Carlo simulation~\cite{deng2003bulk,deng2005surface}, conformal bootstrap~\cite{gliozzi2015boundary}, and fuzzy sphere technique~\cite{zhou2024studying,dedushenko2024ising}, leading to an accurate estimate $\Delta_{\hat \phi} = 1.263 $. 
	With the inclusion of the coupling to $\hat \phi$, the RG equation becomes~\cite{shen2024new}
	\begin{eqnarray}
		\frac{{\rm d}g_1}{{\rm d}l}&=&(2-\Delta_{\hat{\phi}}-K)g_1 \,, \\
		\frac{{\rm d}g_2}{{\rm d}l}&=&(2-4K)g_2\,. 
	\end{eqnarray}
	As $K$ decreases to a critical value $K^\ast = 2 - \Delta_{\hat \phi} \approx 0.736$, $g_1$ becomes relevant and renders a special BKT transition~\footnote{Note that at this point, the two-particle scattering process is still irrelevant.}. 
	This point, according to~\eqref{eq:luttinger}, is approximately $U^\ast_\text{bdy} \approx 3.36$~\footnote{This prediction is not accurate given the renormalization effect from the strong fluctuations in the bulk. We will see the effect in the nonperturbative Monte Carlo calculations.} given the bulk transition occurring at $U_\text{bulk}^\ast \approx 5.65$. 
	Therefore, a complete picture of the boundary criticality for 2D topological insulators described by~\eqref{eq:hamiltonian} emerges: at the bulk 3D Ising critical point, $U_\text{bulk} = U_\text{bulk}^\ast$, the boundary falls into the ordinary universality class, as the boundary Hubbard interaction varies between zero and $U_\text{bdy}^\ast$; as the boundary Hubbard interaction increases to the critical point, $U_\text{bdy}^\ast$, a special BKT transition occurs; exceeding the critical strength, the boundary falls into the extraordinary transition of the 3D Ising universality class, as the edge states are gapped.
	The ordinary and special transitions are distinct from the 3D Ising class due to the enrichment by the helical edge states.
	In particular, the ordinary universality class features a boundary primary field, $S_y$, with a continuously varying critical exponent, $\Delta_S = K$, and the special transition becomes an exotic BKT transition with $\Delta_S^\ast = K^\ast$. 
	On the other hand, the helical edge states give rise to the unique fermionic boundary primary field in the BCFT.
	Its scaling dimension is also a smooth function of the Luttinger parameter before the special transition, $\Delta_F = \frac12 + \frac14 \left( K + \frac1{K} -2 \right)$, which becomes $\Delta_F^\ast = \frac12 + \frac14 \left( K^\ast + \frac1{K^\ast} -2 \right)$ at the special transition.
	
	\begin{figure}
		\centering
		\includegraphics[width=\linewidth]{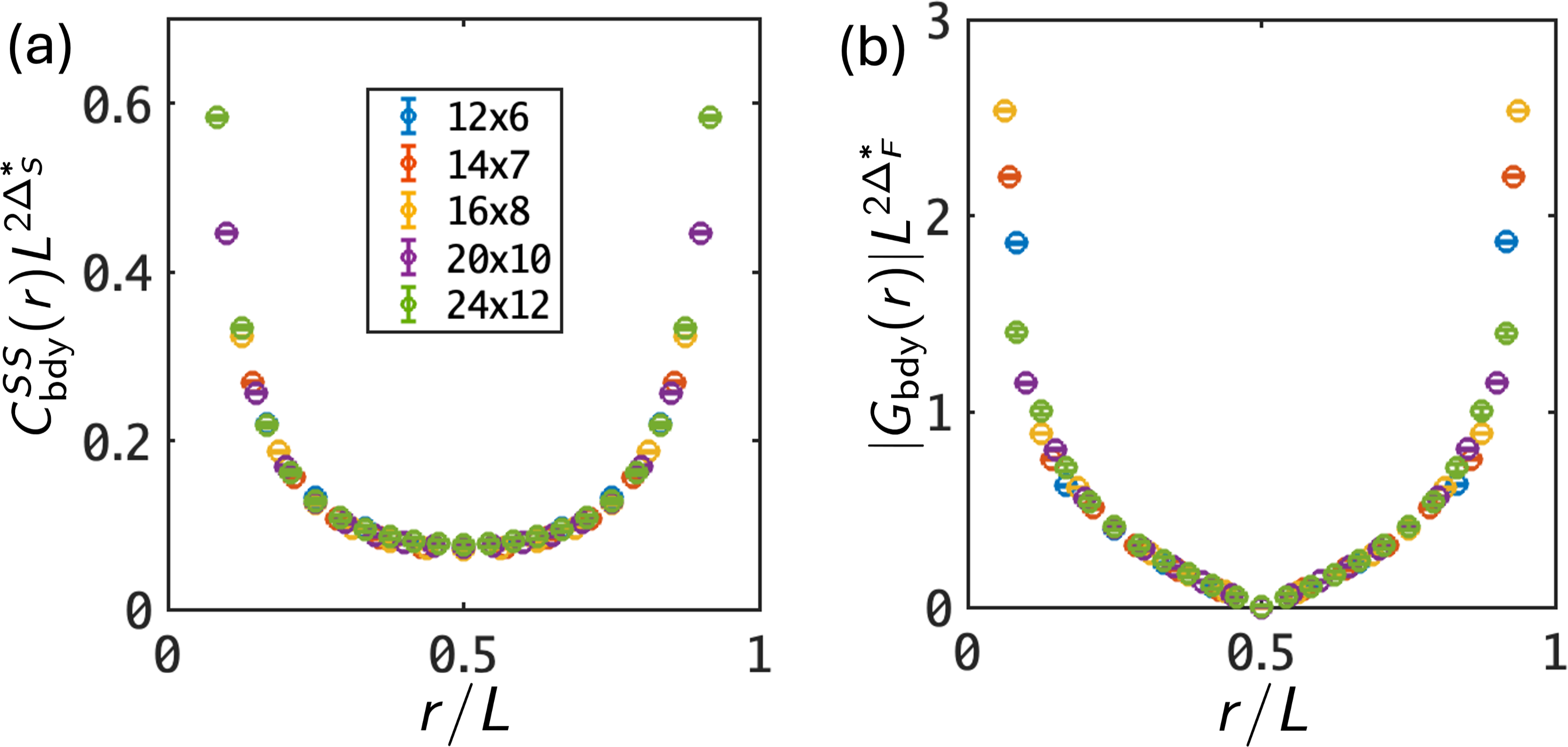}
		\caption{Data collapse at the special transition for (a) the boundary spin-spin correlation function, and (b) the boundary fermion correlation function. 
			The fitted boundary scaling dimensions are given by $\Delta_S^\ast = 0.74$ and $\Delta_F^\ast = 0.48$, respectively.}
		\label{fig:special}
	\end{figure}
	To verify this scenario nonperturbatively, we conducted detailed DQMC simulations at $U^*_\text{bulk}$. 
	The RG-invariant quantity as a function of $U_\text{bdy}$ shown in Fig.~\ref{fig:crossing}(e) exhibits crossings between different system sizes.
	This indicates a special transition.  
	However, the crossings shift as the system size increases. 
	Hence, to accurately determine the special transition point in the thermodynamic limit, an extrapolation is performed in the inset, leading to the special transition $U_\text{bdy}^\ast \approx 1.4$.
	The discrepancy between the $U_\text{bdy}^\ast$ obtained in the simulation and the theoretical prediction is attributed to (i) the  perturbative nature of the analytical method for the Luttinger parameter, and (ii) the presence of gapless fluctuations in the bulk further enhancing the finite-size effect and renormalizing the Luttinger parameter~\cite{shen2024new}. 
	In contrast, when the bulk is gapped for $U_\text{bulk} < U_\text{bulk}^\ast$, the finite-size effect is much smaller. 
	This is seen when $U_\text{bulk} = 5.4 < U_\text{bulk}^\ast$, the crossings show little variation in the left panel of Fig.~\ref{fig:crossing}(d). 
	Conversely, when the bulk is ordered, e.g., at $U_\text{bulk} = 5.8 > U_\text{bulk}^\ast$, the crossings extrapolate to a negative boundary interaction strength $U_\text{bdy}$, indicating an ordered boundary.
	
	Nevertheless, the theoretical prediction of the scaling dimension of the primary spin operator is universal. 
	Indeed, Fig.~\ref{fig:luttinger}(b) shows that, before the special transition, the Luttinger parameter varies continuously with the boundary Hubbard interaction strength. 
	However, the fitted Luttinger parameter deviates from the perturbative analysis, as expected, because the presence of gapless fluctuations in the bulk strongly renormalizes the Luttinger parameter. 
	More crucially, at the special transition extracted from the extrapolation, the Luttinger parameter fitted by the conformal ansatz, as shown in Fig.~\ref{fig:special}, is in full agreement with the theoretical prediction for both the boundary spin operator and the boundary fermion operator. 
	Here, the boundary fermion correlation function is 
	\begin{eqnarray}
		G_\text{bdy}(r) = \frac1{L} \sum_{ \boldsymbol{i} \in \text{bdy} }  \langle c^{\phantom{\dagger}}_{\boldsymbol i, \sigma} c^\dagger_{\boldsymbol{i} + \boldsymbol{r}, \sigma} \rangle.
	\end{eqnarray}
	Notably, the fitted parameters $\Delta_S^\ast = 0.74$ and $\Delta_F^\ast = 0.48$ are perfectly consistent with the RG prediction previously discussed, $\Delta_S^\ast=K^*$ and $\Delta_F^\ast=\frac14(K^*+\frac1{K^*})$, revealing a special transition in interacting topological insulators.

	\paragraph{Concluding remarks}\ssll
	Using DQMC simulations, we present the first nonperturbative study of boundary
	criticality in two-dimensional topological systems with strong electron correlations. 
	The unbiased, large-scale Monte Carlo calculations unveil the enriched boundary criticality due to topological edge states, by tuning the boundary interaction strength in a lattice model with open boundaries. 
	It would be interesting to generalize the method to study other topological systems, such as topological superconductors~\cite{grover2013emergent,li2017edge2} where the special transition falls in a boundary Gross-Neveu-Yukawa universality class~\cite{shen2024new}. 
	Finally, we note that tuning the relative interaction strength between the bulk and the boundary in our model can be potentially realized via edge engineering, e.g., by changing the dielectric environment at the edge~\cite{jia2022tuning}.
	
	\begin{acknowledgments}
		\paragraph{Acknowledgments}\ssll
		We thank Zi-Xiang Li and Zhou-Quan Wan for helpful discussions.
		DQMC simulations were computed using the \textsc{ALF} (Algorithms for Lattice Fermions) package~\cite{assaad2020ALF}. Scaling collapse was aided by \textsc{autoScale.py}~\cite{autoscale}.
		The numerical calculation was performed using high-performance computational resources (HPC) provided by the Louisiana Optical Network Infrastructure.
		Early-stage numerics were performed with the HPC provided by the Information Technology at Tulane University, as well as DARWIN at the University of Delaware, ACES at Texas A\&M University High Performance Research Computing, Anvil at the Rosen Center for Advanced Computing at Purdue University, and Delta system at the University of Illinois Urbana-Champaign and its National Center for Supercomputing Applications, through allocation PHY240119 granted by the Advanced Cyberinfrastructure Coordination Ecosystem: Services \& Support (ACCESS) program, which is supported by National Science Foundation Grants \#2138259, \#2138286, \#2138307, \#2137603, and \#2138296.
		This work is supported in part by a start-up fund (Y.G. and S.-K.J.), the Lavin-Bernick grant (Y.G.) from Tulane University, and the New Cornerstone Science Foundation through the Xplorer Prize (H.Y.). 
	\end{acknowledgments}

	\bibliography{reference.bib}

\begin{thebibliography}{62}%
\makeatletter
\providecommand \@ifxundefined [1]{%
 \@ifx{#1\undefined}
}%
\providecommand \@ifnum [1]{%
 \ifnum #1\expandafter \@firstoftwo
 \else \expandafter \@secondoftwo
 \fi
}%
\providecommand \@ifx [1]{%
 \ifx #1\expandafter \@firstoftwo
 \else \expandafter \@secondoftwo
 \fi
}%
\providecommand \natexlab [1]{#1}%
\providecommand \enquote  [1]{``#1''}%
\providecommand \bibnamefont  [1]{#1}%
\providecommand \bibfnamefont [1]{#1}%
\providecommand \citenamefont [1]{#1}%
\providecommand \href@noop [0]{\@secondoftwo}%
\providecommand \href [0]{\begingroup \@sanitize@url \@href}%
\providecommand \@href[1]{\@@startlink{#1}\@@href}%
\providecommand \@@href[1]{\endgroup#1\@@endlink}%
\providecommand \@sanitize@url [0]{\catcode `\\12\catcode `\$12\catcode
  `\&12\catcode `\#12\catcode `\^12\catcode `\_12\catcode `\%12\relax}%
\providecommand \@@startlink[1]{}%
\providecommand \@@endlink[0]{}%
\providecommand \url  [0]{\begingroup\@sanitize@url \@url }%
\providecommand \@url [1]{\endgroup\@href {#1}{\urlprefix }}%
\providecommand \urlprefix  [0]{URL }%
\providecommand \Eprint [0]{\href }%
\providecommand \doibase [0]{https://doi.org/}%
\providecommand \selectlanguage [0]{\@gobble}%
\providecommand \bibinfo  [0]{\@secondoftwo}%
\providecommand \bibfield  [0]{\@secondoftwo}%
\providecommand \translation [1]{[#1]}%
\providecommand \BibitemOpen [0]{}%
\providecommand \bibitemStop [0]{}%
\providecommand \bibitemNoStop [0]{.\EOS\space}%
\providecommand \EOS [0]{\spacefactor3000\relax}%
\providecommand \BibitemShut  [1]{\csname bibitem#1\endcsname}%
\let\auto@bib@innerbib\@empty
\bibitem [{\citenamefont {Castro~Neto}\ \emph {et~al.}(2009)\citenamefont
  {Castro~Neto}, \citenamefont {Guinea}, \citenamefont {Peres}, \citenamefont
  {Novoselov},\ and\ \citenamefont {Geim}}]{castro2009the}%
  \BibitemOpen
  \bibfield  {author} {\bibinfo {author} {\bibfnamefont {A.~H.}\ \bibnamefont
  {Castro~Neto}}, \bibinfo {author} {\bibfnamefont {F.}~\bibnamefont {Guinea}},
  \bibinfo {author} {\bibfnamefont {N.~M.~R.}\ \bibnamefont {Peres}}, \bibinfo
  {author} {\bibfnamefont {K.~S.}\ \bibnamefont {Novoselov}},\ and\ \bibinfo
  {author} {\bibfnamefont {A.~K.}\ \bibnamefont {Geim}},\ }\bibfield  {title}
  {\bibinfo {title} {The electronic properties of graphene},\ }\href
  {https://doi.org/10.1103/RevModPhys.81.109} {\bibfield  {journal} {\bibinfo
  {journal} {Rev. Mod. Phys.}\ }\textbf {\bibinfo {volume} {81}},\ \bibinfo
  {pages} {109} (\bibinfo {year} {2009})}\BibitemShut {NoStop}%
\bibitem [{\citenamefont {Hasan}\ and\ \citenamefont
  {Kane}(2010)}]{hasan2010colloquium}%
  \BibitemOpen
  \bibfield  {author} {\bibinfo {author} {\bibfnamefont {M.~Z.}\ \bibnamefont
  {Hasan}}\ and\ \bibinfo {author} {\bibfnamefont {C.~L.}\ \bibnamefont
  {Kane}},\ }\bibfield  {title} {\bibinfo {title} {Colloquium: Topological
  insulators},\ }\href {https://doi.org/10.1103/RevModPhys.82.3045} {\bibfield
  {journal} {\bibinfo  {journal} {Rev. Mod. Phys.}\ }\textbf {\bibinfo {volume}
  {82}},\ \bibinfo {pages} {3045} (\bibinfo {year} {2010})}\BibitemShut
  {NoStop}%
\bibitem [{\citenamefont {Qi}\ and\ \citenamefont
  {Zhang}(2011)}]{qi2011topological}%
  \BibitemOpen
  \bibfield  {author} {\bibinfo {author} {\bibfnamefont {X.-L.}\ \bibnamefont
  {Qi}}\ and\ \bibinfo {author} {\bibfnamefont {S.-C.}\ \bibnamefont {Zhang}},\
  }\bibfield  {title} {\bibinfo {title} {Topological insulators and
  superconductors},\ }\href {https://doi.org/10.1103/RevModPhys.83.1057}
  {\bibfield  {journal} {\bibinfo  {journal} {Rev. Mod. Phys.}\ }\textbf
  {\bibinfo {volume} {83}},\ \bibinfo {pages} {1057} (\bibinfo {year}
  {2011})}\BibitemShut {NoStop}%
\bibitem [{\citenamefont {Armitage}\ \emph {et~al.}(2018)\citenamefont
  {Armitage}, \citenamefont {Mele},\ and\ \citenamefont
  {Vishwanath}}]{armitage2018weyl}%
  \BibitemOpen
  \bibfield  {author} {\bibinfo {author} {\bibfnamefont {N.~P.}\ \bibnamefont
  {Armitage}}, \bibinfo {author} {\bibfnamefont {E.~J.}\ \bibnamefont {Mele}},\
  and\ \bibinfo {author} {\bibfnamefont {A.}~\bibnamefont {Vishwanath}},\
  }\bibfield  {title} {\bibinfo {title} {{Weyl} and {Dirac} semimetals in
  three-dimensional solids},\ }\href
  {https://doi.org/10.1103/RevModPhys.90.015001} {\bibfield  {journal}
  {\bibinfo  {journal} {Rev. Mod. Phys.}\ }\textbf {\bibinfo {volume} {90}},\
  \bibinfo {pages} {015001} (\bibinfo {year} {2018})}\BibitemShut {NoStop}%
\bibitem [{\citenamefont {Lv}\ \emph {et~al.}(2021)\citenamefont {Lv},
  \citenamefont {Qian},\ and\ \citenamefont {Ding}}]{lv2021experimental}%
  \BibitemOpen
  \bibfield  {author} {\bibinfo {author} {\bibfnamefont {B.~Q.}\ \bibnamefont
  {Lv}}, \bibinfo {author} {\bibfnamefont {T.}~\bibnamefont {Qian}},\ and\
  \bibinfo {author} {\bibfnamefont {H.}~\bibnamefont {Ding}},\ }\bibfield
  {title} {\bibinfo {title} {Experimental perspective on three-dimensional
  topological semimetals},\ }\href
  {https://doi.org/10.1103/RevModPhys.93.025002} {\bibfield  {journal}
  {\bibinfo  {journal} {Rev. Mod. Phys.}\ }\textbf {\bibinfo {volume} {93}},\
  \bibinfo {pages} {025002} (\bibinfo {year} {2021})}\BibitemShut {NoStop}%
\bibitem [{\citenamefont {Klitzing}\ \emph {et~al.}(1980)\citenamefont
  {Klitzing}, \citenamefont {Dorda},\ and\ \citenamefont
  {Pepper}}]{klitzing1980new}%
  \BibitemOpen
  \bibfield  {author} {\bibinfo {author} {\bibfnamefont {K.~v.}\ \bibnamefont
  {Klitzing}}, \bibinfo {author} {\bibfnamefont {G.}~\bibnamefont {Dorda}},\
  and\ \bibinfo {author} {\bibfnamefont {M.}~\bibnamefont {Pepper}},\
  }\bibfield  {title} {\bibinfo {title} {New method for high-accuracy
  determination of the fine-structure constant based on quantized {Hall}
  resistance},\ }\href {https://doi.org/10.1103/PhysRevLett.45.494} {\bibfield
  {journal} {\bibinfo  {journal} {Phys. Rev. Lett.}\ }\textbf {\bibinfo
  {volume} {45}},\ \bibinfo {pages} {494} (\bibinfo {year} {1980})}\BibitemShut
  {NoStop}%
\bibitem [{\citenamefont {Laughlin}(1981)}]{laughlin1981quantized}%
  \BibitemOpen
  \bibfield  {author} {\bibinfo {author} {\bibfnamefont {R.~B.}\ \bibnamefont
  {Laughlin}},\ }\bibfield  {title} {\bibinfo {title} {Quantized {Hall}
  conductivity in two dimensions},\ }\href
  {https://doi.org/10.1103/PhysRevB.23.5632} {\bibfield  {journal} {\bibinfo
  {journal} {Phys. Rev. B}\ }\textbf {\bibinfo {volume} {23}},\ \bibinfo
  {pages} {5632} (\bibinfo {year} {1981})}\BibitemShut {NoStop}%
\bibitem [{\citenamefont {Tsui}\ \emph {et~al.}(1982)\citenamefont {Tsui},
  \citenamefont {Stormer},\ and\ \citenamefont {Gossard}}]{tsui1982two}%
  \BibitemOpen
  \bibfield  {author} {\bibinfo {author} {\bibfnamefont {D.~C.}\ \bibnamefont
  {Tsui}}, \bibinfo {author} {\bibfnamefont {H.~L.}\ \bibnamefont {Stormer}},\
  and\ \bibinfo {author} {\bibfnamefont {A.~C.}\ \bibnamefont {Gossard}},\
  }\bibfield  {title} {\bibinfo {title} {Two-dimensional magnetotransport in
  the extreme quantum limit},\ }\href
  {https://doi.org/10.1103/PhysRevLett.48.1559} {\bibfield  {journal} {\bibinfo
   {journal} {Phys. Rev. Lett.}\ }\textbf {\bibinfo {volume} {48}},\ \bibinfo
  {pages} {1559} (\bibinfo {year} {1982})}\BibitemShut {NoStop}%
\bibitem [{\citenamefont {Kane}\ and\ \citenamefont
  {Mele}(2005{\natexlab{a}})}]{kane2005quantum}%
  \BibitemOpen
  \bibfield  {author} {\bibinfo {author} {\bibfnamefont {C.~L.}\ \bibnamefont
  {Kane}}\ and\ \bibinfo {author} {\bibfnamefont {E.~J.}\ \bibnamefont
  {Mele}},\ }\bibfield  {title} {\bibinfo {title} {Quantum spin {Hall} effect
  in graphene},\ }\href {https://doi.org/10.1103/PhysRevLett.95.226801}
  {\bibfield  {journal} {\bibinfo  {journal} {Phys. Rev. Lett.}\ }\textbf
  {\bibinfo {volume} {95}},\ \bibinfo {pages} {226801} (\bibinfo {year}
  {2005}{\natexlab{a}})}\BibitemShut {NoStop}%
\bibitem [{\citenamefont {Bernevig}\ \emph {et~al.}(2006)\citenamefont
  {Bernevig}, \citenamefont {Hughes},\ and\ \citenamefont
  {Zhang}}]{bernevig2006quantum}%
  \BibitemOpen
  \bibfield  {author} {\bibinfo {author} {\bibfnamefont {B.~A.}\ \bibnamefont
  {Bernevig}}, \bibinfo {author} {\bibfnamefont {T.~L.}\ \bibnamefont
  {Hughes}},\ and\ \bibinfo {author} {\bibfnamefont {S.-C.}\ \bibnamefont
  {Zhang}},\ }\bibfield  {title} {\bibinfo {title} {Quantum spin {Hall} effect
  and topological phase transition in {HgTe} quantum wells},\ }\href
  {https://doi.org/10.1126/science.1133734} {\bibfield  {journal} {\bibinfo
  {journal} {Science}\ }\textbf {\bibinfo {volume} {314}},\ \bibinfo {pages}
  {1757} (\bibinfo {year} {2006})}\BibitemShut {NoStop}%
\bibitem [{\citenamefont {König}\ \emph {et~al.}(2007)\citenamefont {König},
  \citenamefont {Wiedmann}, \citenamefont {Brüne}, \citenamefont {Roth},
  \citenamefont {Buhmann}, \citenamefont {Molenkamp}, \citenamefont {Qi},\ and\
  \citenamefont {Zhang}}]{konig2007quantum}%
  \BibitemOpen
  \bibfield  {author} {\bibinfo {author} {\bibfnamefont {M.}~\bibnamefont
  {König}}, \bibinfo {author} {\bibfnamefont {S.}~\bibnamefont {Wiedmann}},
  \bibinfo {author} {\bibfnamefont {C.}~\bibnamefont {Brüne}}, \bibinfo
  {author} {\bibfnamefont {A.}~\bibnamefont {Roth}}, \bibinfo {author}
  {\bibfnamefont {H.}~\bibnamefont {Buhmann}}, \bibinfo {author} {\bibfnamefont
  {L.~W.}\ \bibnamefont {Molenkamp}}, \bibinfo {author} {\bibfnamefont {X.-L.}\
  \bibnamefont {Qi}},\ and\ \bibinfo {author} {\bibfnamefont {S.-C.}\
  \bibnamefont {Zhang}},\ }\bibfield  {title} {\bibinfo {title} {Quantum spin
  {Hall} insulator state in {HgTe} quantum wells},\ }\href
  {https://doi.org/10.1126/science.1148047} {\bibfield  {journal} {\bibinfo
  {journal} {Science}\ }\textbf {\bibinfo {volume} {318}},\ \bibinfo {pages}
  {766} (\bibinfo {year} {2007})}\BibitemShut {NoStop}%
\bibitem [{\citenamefont {Schnyder}\ \emph {et~al.}(2008)\citenamefont
  {Schnyder}, \citenamefont {Ryu}, \citenamefont {Furusaki},\ and\
  \citenamefont {Ludwig}}]{schnyder2008classification}%
  \BibitemOpen
  \bibfield  {author} {\bibinfo {author} {\bibfnamefont {A.~P.}\ \bibnamefont
  {Schnyder}}, \bibinfo {author} {\bibfnamefont {S.}~\bibnamefont {Ryu}},
  \bibinfo {author} {\bibfnamefont {A.}~\bibnamefont {Furusaki}},\ and\
  \bibinfo {author} {\bibfnamefont {A.~W.~W.}\ \bibnamefont {Ludwig}},\
  }\bibfield  {title} {\bibinfo {title} {Classification of topological
  insulators and superconductors in three spatial dimensions},\ }\href
  {https://doi.org/10.1103/PhysRevB.78.195125} {\bibfield  {journal} {\bibinfo
  {journal} {Phys. Rev. B}\ }\textbf {\bibinfo {volume} {78}},\ \bibinfo
  {pages} {195125} (\bibinfo {year} {2008})}\BibitemShut {NoStop}%
\bibitem [{\citenamefont {Kitaev}(2009)}]{kitaev2009periodic}%
  \BibitemOpen
  \bibfield  {author} {\bibinfo {author} {\bibfnamefont {A.}~\bibnamefont
  {Kitaev}},\ }\bibfield  {title} {\bibinfo {title} {{Periodic table for
  topological insulators and superconductors}},\ }\href
  {https://doi.org/10.1063/1.3149495} {\bibfield  {journal} {\bibinfo
  {journal} {AIP Conf. Proc.}\ }\textbf {\bibinfo {volume} {1134}},\ \bibinfo
  {pages} {22} (\bibinfo {year} {2009})},\ \Eprint
  {https://arxiv.org/abs/0901.2686} {arXiv:0901.2686 [cond-mat.mes-hall]}
  \BibitemShut {NoStop}%
\bibitem [{\citenamefont {Sato}\ and\ \citenamefont
  {Ando}(2017)}]{sato2017topological}%
  \BibitemOpen
  \bibfield  {author} {\bibinfo {author} {\bibfnamefont {M.}~\bibnamefont
  {Sato}}\ and\ \bibinfo {author} {\bibfnamefont {Y.}~\bibnamefont {Ando}},\
  }\bibfield  {title} {\bibinfo {title} {Topological superconductors: a
  review},\ }\href {https://doi.org/10.1088/1361-6633/aa6ac7} {\bibfield
  {journal} {\bibinfo  {journal} {Rep. Prog. Phys.}\ }\textbf {\bibinfo
  {volume} {80}},\ \bibinfo {pages} {076501} (\bibinfo {year}
  {2017})}\BibitemShut {NoStop}%
\bibitem [{\citenamefont {Kane}\ and\ \citenamefont
  {Mele}(2005{\natexlab{b}})}]{kane2005Z2}%
  \BibitemOpen
  \bibfield  {author} {\bibinfo {author} {\bibfnamefont {C.~L.}\ \bibnamefont
  {Kane}}\ and\ \bibinfo {author} {\bibfnamefont {E.~J.}\ \bibnamefont
  {Mele}},\ }\bibfield  {title} {\bibinfo {title} {${Z}_{2}$ topological order
  and the quantum spin {Hall} effect},\ }\href
  {https://doi.org/10.1103/PhysRevLett.95.146802} {\bibfield  {journal}
  {\bibinfo  {journal} {Phys. Rev. Lett.}\ }\textbf {\bibinfo {volume} {95}},\
  \bibinfo {pages} {146802} (\bibinfo {year} {2005}{\natexlab{b}})}\BibitemShut
  {NoStop}%
\bibitem [{\citenamefont {Wu}\ \emph {et~al.}(2006)\citenamefont {Wu},
  \citenamefont {Bernevig},\ and\ \citenamefont {Zhang}}]{wu2006helical}%
  \BibitemOpen
  \bibfield  {author} {\bibinfo {author} {\bibfnamefont {C.}~\bibnamefont
  {Wu}}, \bibinfo {author} {\bibfnamefont {B.~A.}\ \bibnamefont {Bernevig}},\
  and\ \bibinfo {author} {\bibfnamefont {S.-C.}\ \bibnamefont {Zhang}},\
  }\bibfield  {title} {\bibinfo {title} {Helical liquid and the edge of quantum
  spin {Hall} systems},\ }\href {https://doi.org/10.1103/PhysRevLett.96.106401}
  {\bibfield  {journal} {\bibinfo  {journal} {Phys. Rev. Lett.}\ }\textbf
  {\bibinfo {volume} {96}},\ \bibinfo {pages} {106401} (\bibinfo {year}
  {2006})}\BibitemShut {NoStop}%
\bibitem [{\citenamefont {Xu}\ and\ \citenamefont {Moore}(2006)}]{CKXu2006PRB}%
  \BibitemOpen
  \bibfield  {author} {\bibinfo {author} {\bibfnamefont {C.}~\bibnamefont
  {Xu}}\ and\ \bibinfo {author} {\bibfnamefont {J.~E.}\ \bibnamefont {Moore}},\
  }\bibfield  {title} {\bibinfo {title} {Stability of the quantum spin {Hall}
  effect: Effects of interactions, disorder, and ${\mathbb{z}}_{2}$ topology},\
  }\href {https://doi.org/10.1103/PhysRevB.73.045322} {\bibfield  {journal}
  {\bibinfo  {journal} {Phys. Rev. B}\ }\textbf {\bibinfo {volume} {73}},\
  \bibinfo {pages} {045322} (\bibinfo {year} {2006})}\BibitemShut {NoStop}%
\bibitem [{\citenamefont {Hsu}\ \emph {et~al.}(2021)\citenamefont {Hsu},
  \citenamefont {Stano}, \citenamefont {Klinovaja},\ and\ \citenamefont
  {Loss}}]{hsu2021helical}%
  \BibitemOpen
  \bibfield  {author} {\bibinfo {author} {\bibfnamefont {C.-H.}\ \bibnamefont
  {Hsu}}, \bibinfo {author} {\bibfnamefont {P.}~\bibnamefont {Stano}}, \bibinfo
  {author} {\bibfnamefont {J.}~\bibnamefont {Klinovaja}},\ and\ \bibinfo
  {author} {\bibfnamefont {D.}~\bibnamefont {Loss}},\ }\bibfield  {title}
  {\bibinfo {title} {Helical liquids in semiconductors},\ }\href@noop {}
  {\bibfield  {journal} {\bibinfo  {journal} {Semiconductor Science and
  Technology}\ }\textbf {\bibinfo {volume} {36}},\ \bibinfo {pages} {123003}
  (\bibinfo {year} {2021})}\BibitemShut {NoStop}%
\bibitem [{\citenamefont {Cardy}(1984)}]{cardy1984conformal}%
  \BibitemOpen
  \bibfield  {author} {\bibinfo {author} {\bibfnamefont {J.~L.}\ \bibnamefont
  {Cardy}},\ }\bibfield  {title} {\bibinfo {title} {Conformal invariance and
  surface critical behavior},\ }\href
  {https://doi.org/https://doi.org/10.1016/0550-3213(84)90241-4} {\bibfield
  {journal} {\bibinfo  {journal} {Nucl. Phys. B}\ }\textbf {\bibinfo {volume}
  {240}},\ \bibinfo {pages} {514} (\bibinfo {year} {1984})}\BibitemShut
  {NoStop}%
\bibitem [{\citenamefont {Burkhardt}\ and\ \citenamefont
  {Diehl}(1994)}]{burkhardt1994ordinary}%
  \BibitemOpen
  \bibfield  {author} {\bibinfo {author} {\bibfnamefont {T.~W.}\ \bibnamefont
  {Burkhardt}}\ and\ \bibinfo {author} {\bibfnamefont {H.~W.}\ \bibnamefont
  {Diehl}},\ }\bibfield  {title} {\bibinfo {title} {Ordinary, extraordinary,
  and normal surface transitions: Extraordinary-normal equivalence and simple
  explanation of
  ${|T\ensuremath{-}{T}_{c}|}^{2\ensuremath{-}\ensuremath{\alpha}}$
  singularities},\ }\href {https://doi.org/10.1103/PhysRevB.50.3894} {\bibfield
   {journal} {\bibinfo  {journal} {Phys. Rev. B}\ }\textbf {\bibinfo {volume}
  {50}},\ \bibinfo {pages} {3894} (\bibinfo {year} {1994})}\BibitemShut
  {NoStop}%
\bibitem [{\citenamefont {Liendo}\ \emph {et~al.}(2013)\citenamefont {Liendo},
  \citenamefont {Rastelli},\ and\ \citenamefont {van
  Rees}}]{liendo2013bootstrap}%
  \BibitemOpen
  \bibfield  {author} {\bibinfo {author} {\bibfnamefont {P.}~\bibnamefont
  {Liendo}}, \bibinfo {author} {\bibfnamefont {L.}~\bibnamefont {Rastelli}},\
  and\ \bibinfo {author} {\bibfnamefont {B.~C.}\ \bibnamefont {van Rees}},\
  }\bibfield  {title} {\bibinfo {title} {The bootstrap program for boundary
  {CFT$_d$}},\ }\href {https://doi.org/10.1007/JHEP07(2013)113} {\bibfield
  {journal} {\bibinfo  {journal} {JHEP}\ }\textbf {\bibinfo {volume} {07}},\
  \bibinfo {pages} {113}},\ \Eprint {https://arxiv.org/abs/1210.4258}
  {arXiv:1210.4258 [hep-th]} \BibitemShut {NoStop}%
\bibitem [{\citenamefont {Andrei}\ \emph {et~al.}(2020)\citenamefont {Andrei},
  \citenamefont {Bissi}, \citenamefont {Buican}, \citenamefont {Cardy},
  \citenamefont {Dorey}, \citenamefont {Drukker}, \citenamefont {Erdmenger},
  \citenamefont {Friedan}, \citenamefont {Fursaev}, \citenamefont {Konechny},
  \citenamefont {Kristjansen}, \citenamefont {Makabe}, \citenamefont
  {Nakayama}, \citenamefont {O’Bannon}, \citenamefont {Parini}, \citenamefont
  {Robinson}, \citenamefont {Ryu}, \citenamefont {Schmidt-Colinet},
  \citenamefont {Schomerus}, \citenamefont {Schweigert},\ and\ \citenamefont
  {Watts}}]{andrei2018boundary}%
  \BibitemOpen
  \bibfield  {author} {\bibinfo {author} {\bibfnamefont {N.}~\bibnamefont
  {Andrei}}, \bibinfo {author} {\bibfnamefont {A.}~\bibnamefont {Bissi}},
  \bibinfo {author} {\bibfnamefont {M.}~\bibnamefont {Buican}}, \bibinfo
  {author} {\bibfnamefont {J.}~\bibnamefont {Cardy}}, \bibinfo {author}
  {\bibfnamefont {P.}~\bibnamefont {Dorey}}, \bibinfo {author} {\bibfnamefont
  {N.}~\bibnamefont {Drukker}}, \bibinfo {author} {\bibfnamefont
  {J.}~\bibnamefont {Erdmenger}}, \bibinfo {author} {\bibfnamefont
  {D.}~\bibnamefont {Friedan}}, \bibinfo {author} {\bibfnamefont
  {D.}~\bibnamefont {Fursaev}}, \bibinfo {author} {\bibfnamefont
  {A.}~\bibnamefont {Konechny}}, \bibinfo {author} {\bibfnamefont
  {C.}~\bibnamefont {Kristjansen}}, \bibinfo {author} {\bibfnamefont
  {I.}~\bibnamefont {Makabe}}, \bibinfo {author} {\bibfnamefont
  {Y.}~\bibnamefont {Nakayama}}, \bibinfo {author} {\bibfnamefont
  {A.}~\bibnamefont {O’Bannon}}, \bibinfo {author} {\bibfnamefont
  {R.}~\bibnamefont {Parini}}, \bibinfo {author} {\bibfnamefont
  {B.}~\bibnamefont {Robinson}}, \bibinfo {author} {\bibfnamefont
  {S.}~\bibnamefont {Ryu}}, \bibinfo {author} {\bibfnamefont {C.}~\bibnamefont
  {Schmidt-Colinet}}, \bibinfo {author} {\bibfnamefont {V.}~\bibnamefont
  {Schomerus}}, \bibinfo {author} {\bibfnamefont {C.}~\bibnamefont
  {Schweigert}},\ and\ \bibinfo {author} {\bibfnamefont {G.~M.~T.}\
  \bibnamefont {Watts}},\ }\bibfield  {title} {\bibinfo {title} {Boundary and
  defect {CFT}: Open problems and applications},\ }\href
  {https://doi.org/10.1088/1751-8121/abb0fe} {\bibfield  {journal} {\bibinfo
  {journal} {J. Phys. A}\ }\textbf {\bibinfo {volume} {53}},\ \bibinfo {pages}
  {453002} (\bibinfo {year} {2020})},\ \Eprint
  {https://arxiv.org/abs/1810.05697} {arXiv:1810.05697 [hep-th]} \BibitemShut
  {NoStop}%
\bibitem [{\citenamefont {Domb}\ and\ \citenamefont
  {Lebowitz}(1986)}]{domb1986phase}%
  \BibitemOpen
  \bibinfo {editor} {\bibfnamefont {C.}~\bibnamefont {Domb}}\ and\ \bibinfo
  {editor} {\bibfnamefont {J.~L.}\ \bibnamefont {Lebowitz}},\ eds.,\ \href@noop
  {} {\emph {\bibinfo {title} {Phase Transitions and Critical Phenomena}}},\
  Vol.~\bibinfo {volume} {10}\ (\bibinfo  {publisher} {Academic Press},\
  \bibinfo {address} {London},\ \bibinfo {year} {1986})\BibitemShut {NoStop}%
\bibitem [{\citenamefont {Diehl}(1997)}]{diehl1996the}%
  \BibitemOpen
  \bibfield  {author} {\bibinfo {author} {\bibfnamefont {H.~W.}\ \bibnamefont
  {Diehl}},\ }\bibfield  {title} {\bibinfo {title} {{The theory of boundary
  critical phenomena}},\ }\href {https://doi.org/10.1142/S0217979297001751}
  {\bibfield  {journal} {\bibinfo  {journal} {Int. J. Mod. Phys. B}\ }\textbf
  {\bibinfo {volume} {11}},\ \bibinfo {pages} {3503} (\bibinfo {year}
  {1997})},\ \Eprint {https://arxiv.org/abs/cond-mat/9610143}
  {arXiv:cond-mat/9610143} \BibitemShut {NoStop}%
\bibitem [{\citenamefont {Cardy}(1996)}]{cardy1996scaling}%
  \BibitemOpen
  \bibfield  {author} {\bibinfo {author} {\bibfnamefont {J.}~\bibnamefont
  {Cardy}},\ }\href {https://doi.org/10.1017/CBO9781316036440} {\emph {\bibinfo
  {title} {Scaling and Renormalization in Statistical Physics}}},\ Cambridge
  Lecture Notes in Physics\ (\bibinfo  {publisher} {Cambridge University
  Press},\ \bibinfo {address} {Cambridge},\ \bibinfo {year} {1996})\BibitemShut
  {NoStop}%
\bibitem [{\citenamefont {Scaffidi}\ \emph {et~al.}(2017)\citenamefont
  {Scaffidi}, \citenamefont {Parker},\ and\ \citenamefont
  {Vasseur}}]{scaffidi2017gapless}%
  \BibitemOpen
  \bibfield  {author} {\bibinfo {author} {\bibfnamefont {T.}~\bibnamefont
  {Scaffidi}}, \bibinfo {author} {\bibfnamefont {D.~E.}\ \bibnamefont
  {Parker}},\ and\ \bibinfo {author} {\bibfnamefont {R.}~\bibnamefont
  {Vasseur}},\ }\bibfield  {title} {\bibinfo {title} {Gapless
  symmetry-protected topological order},\ }\href
  {https://doi.org/10.1103/PhysRevX.7.041048} {\bibfield  {journal} {\bibinfo
  {journal} {Phys. Rev. X}\ }\textbf {\bibinfo {volume} {7}},\ \bibinfo {pages}
  {041048} (\bibinfo {year} {2017})}\BibitemShut {NoStop}%
\bibitem [{\citenamefont {Zhang}\ and\ \citenamefont
  {Wang}(2017)}]{zhang2017unconventional}%
  \BibitemOpen
  \bibfield  {author} {\bibinfo {author} {\bibfnamefont {L.}~\bibnamefont
  {Zhang}}\ and\ \bibinfo {author} {\bibfnamefont {F.}~\bibnamefont {Wang}},\
  }\bibfield  {title} {\bibinfo {title} {Unconventional surface critical
  behavior induced by a quantum phase transition from the two-dimensional
  {Affleck}-{Kennedy}-{Lieb}-{Tasaki} phase to a {N\'eel}-ordered phase},\
  }\href {https://doi.org/10.1103/PhysRevLett.118.087201} {\bibfield  {journal}
  {\bibinfo  {journal} {Phys. Rev. Lett.}\ }\textbf {\bibinfo {volume} {118}},\
  \bibinfo {pages} {087201} (\bibinfo {year} {2017})}\BibitemShut {NoStop}%
\bibitem [{\citenamefont {Wu}\ \emph {et~al.}(2020)\citenamefont {Wu},
  \citenamefont {Xu}, \citenamefont {Geng}, \citenamefont {Jian},\ and\
  \citenamefont {Xu}}]{wu2020boundary}%
  \BibitemOpen
  \bibfield  {author} {\bibinfo {author} {\bibfnamefont {X.-C.}\ \bibnamefont
  {Wu}}, \bibinfo {author} {\bibfnamefont {Y.}~\bibnamefont {Xu}}, \bibinfo
  {author} {\bibfnamefont {H.}~\bibnamefont {Geng}}, \bibinfo {author}
  {\bibfnamefont {C.-M.}\ \bibnamefont {Jian}},\ and\ \bibinfo {author}
  {\bibfnamefont {C.}~\bibnamefont {Xu}},\ }\bibfield  {title} {\bibinfo
  {title} {Boundary criticality of topological quantum phase transitions in
  two-dimensional systems},\ }\href
  {https://doi.org/10.1103/PhysRevB.101.174406} {\bibfield  {journal} {\bibinfo
   {journal} {Phys. Rev. B}\ }\textbf {\bibinfo {volume} {101}},\ \bibinfo
  {pages} {174406} (\bibinfo {year} {2020})}\BibitemShut {NoStop}%
\bibitem [{\citenamefont {Verresen}\ \emph {et~al.}(2021)\citenamefont
  {Verresen}, \citenamefont {Thorngren}, \citenamefont {Jones},\ and\
  \citenamefont {Pollmann}}]{verresen2021gapless}%
  \BibitemOpen
  \bibfield  {author} {\bibinfo {author} {\bibfnamefont {R.}~\bibnamefont
  {Verresen}}, \bibinfo {author} {\bibfnamefont {R.}~\bibnamefont {Thorngren}},
  \bibinfo {author} {\bibfnamefont {N.~G.}\ \bibnamefont {Jones}},\ and\
  \bibinfo {author} {\bibfnamefont {F.}~\bibnamefont {Pollmann}},\ }\bibfield
  {title} {\bibinfo {title} {Gapless topological phases and symmetry-enriched
  quantum criticality},\ }\href {https://doi.org/10.1103/PhysRevX.11.041059}
  {\bibfield  {journal} {\bibinfo  {journal} {Phys. Rev. X}\ }\textbf {\bibinfo
  {volume} {11}},\ \bibinfo {pages} {041059} (\bibinfo {year}
  {2021})}\BibitemShut {NoStop}%
\bibitem [{\citenamefont {Ma}\ \emph {et~al.}(2022)\citenamefont {Ma},
  \citenamefont {Zou},\ and\ \citenamefont {Wang}}]{ma2022edge}%
  \BibitemOpen
  \bibfield  {author} {\bibinfo {author} {\bibfnamefont {R.}~\bibnamefont
  {Ma}}, \bibinfo {author} {\bibfnamefont {L.}~\bibnamefont {Zou}},\ and\
  \bibinfo {author} {\bibfnamefont {C.}~\bibnamefont {Wang}},\ }\bibfield
  {title} {\bibinfo {title} {{Edge physics at the deconfined transition between
  a quantum spin {Hall} insulator and a superconductor}},\ }\href
  {https://doi.org/10.21468/SciPostPhys.12.6.196} {\bibfield  {journal}
  {\bibinfo  {journal} {SciPost Phys.}\ }\textbf {\bibinfo {volume} {12}},\
  \bibinfo {pages} {196} (\bibinfo {year} {2022})}\BibitemShut {NoStop}%
\bibitem [{\citenamefont {Yu}\ \emph {et~al.}(2022)\citenamefont {Yu},
  \citenamefont {Huang}, \citenamefont {Song}, \citenamefont {Xu},
  \citenamefont {Ding},\ and\ \citenamefont {Zhang}}]{yu2022conformal}%
  \BibitemOpen
  \bibfield  {author} {\bibinfo {author} {\bibfnamefont {X.-J.}\ \bibnamefont
  {Yu}}, \bibinfo {author} {\bibfnamefont {R.-Z.}\ \bibnamefont {Huang}},
  \bibinfo {author} {\bibfnamefont {H.-H.}\ \bibnamefont {Song}}, \bibinfo
  {author} {\bibfnamefont {L.}~\bibnamefont {Xu}}, \bibinfo {author}
  {\bibfnamefont {C.}~\bibnamefont {Ding}},\ and\ \bibinfo {author}
  {\bibfnamefont {L.}~\bibnamefont {Zhang}},\ }\bibfield  {title} {\bibinfo
  {title} {Conformal boundary conditions of symmetry-enriched quantum critical
  spin chains},\ }\href {https://doi.org/10.1103/PhysRevLett.129.210601}
  {\bibfield  {journal} {\bibinfo  {journal} {Phys. Rev. Lett.}\ }\textbf
  {\bibinfo {volume} {129}},\ \bibinfo {pages} {210601} (\bibinfo {year}
  {2022})}\BibitemShut {NoStop}%
\bibitem [{\citenamefont {Yu}\ \emph {et~al.}(2024)\citenamefont {Yu},
  \citenamefont {Yang}, \citenamefont {Lin},\ and\ \citenamefont
  {Jian}}]{yu2024universal}%
  \BibitemOpen
  \bibfield  {author} {\bibinfo {author} {\bibfnamefont {X.-J.}\ \bibnamefont
  {Yu}}, \bibinfo {author} {\bibfnamefont {S.}~\bibnamefont {Yang}}, \bibinfo
  {author} {\bibfnamefont {H.-Q.}\ \bibnamefont {Lin}},\ and\ \bibinfo {author}
  {\bibfnamefont {S.-K.}\ \bibnamefont {Jian}},\ }\bibfield  {title} {\bibinfo
  {title} {Universal entanglement spectrum in one-dimensional gapless symmetry
  protected topological states},\ }\href
  {https://doi.org/10.1103/PhysRevLett.133.026601} {\bibfield  {journal}
  {\bibinfo  {journal} {Phys. Rev. Lett.}\ }\textbf {\bibinfo {volume} {133}},\
  \bibinfo {pages} {026601} (\bibinfo {year} {2024})}\BibitemShut {NoStop}%
\bibitem [{\citenamefont {Shen}\ \emph {et~al.}(2025)\citenamefont {Shen},
  \citenamefont {Wu},\ and\ \citenamefont {Jian}}]{shen2024new}%
  \BibitemOpen
  \bibfield  {author} {\bibinfo {author} {\bibfnamefont {X.}~\bibnamefont
  {Shen}}, \bibinfo {author} {\bibfnamefont {Z.}~\bibnamefont {Wu}},\ and\
  \bibinfo {author} {\bibfnamefont {S.-K.}\ \bibnamefont {Jian}},\ }\bibfield
  {title} {\bibinfo {title} {Boundary and defect criticality in topological
  insulators and superconductors},\ }\href {https://doi.org/10.1103/4lv4-mc81}
  {\bibfield  {journal} {\bibinfo  {journal} {Phys. Rev. B}\ }\textbf {\bibinfo
  {volume} {112}},\ \bibinfo {pages} {L041118} (\bibinfo {year}
  {2025})}\BibitemShut {NoStop}%
\bibitem [{\citenamefont {Fehske}\ \emph {et~al.}(2007)\citenamefont {Fehske},
  \citenamefont {Schneider},\ and\ \citenamefont
  {Weisse}}]{fehske2007computational}%
  \BibitemOpen
  \bibfield  {author} {\bibinfo {author} {\bibfnamefont {H.}~\bibnamefont
  {Fehske}}, \bibinfo {author} {\bibfnamefont {R.}~\bibnamefont {Schneider}},\
  and\ \bibinfo {author} {\bibfnamefont {A.}~\bibnamefont {Weisse}},\
  }\href@noop {} {\emph {\bibinfo {title} {Computational many-particle
  physics}}},\ Vol.\ \bibinfo {volume} {739}\ (\bibinfo  {publisher}
  {Springer},\ \bibinfo {address} {Berlin, Heidelberg},\ \bibinfo {year}
  {2007})\BibitemShut {NoStop}%
\bibitem [{\citenamefont {Assaad}\ \emph {et~al.}(2022)\citenamefont {Assaad},
  \citenamefont {Bercx}, \citenamefont {Goth}, \citenamefont {G\"otz},
  \citenamefont {Hofmann}, \citenamefont {Huffman}, \citenamefont {Liu},
  \citenamefont {Parisen~Toldin}, \citenamefont {Portela},\ and\ \citenamefont
  {Schwab}}]{assaad2020ALF}%
  \BibitemOpen
  \bibfield  {author} {\bibinfo {author} {\bibfnamefont {F.~F.}\ \bibnamefont
  {Assaad}}, \bibinfo {author} {\bibfnamefont {M.}~\bibnamefont {Bercx}},
  \bibinfo {author} {\bibfnamefont {F.}~\bibnamefont {Goth}}, \bibinfo {author}
  {\bibfnamefont {A.}~\bibnamefont {G\"otz}}, \bibinfo {author} {\bibfnamefont
  {J.~S.}\ \bibnamefont {Hofmann}}, \bibinfo {author} {\bibfnamefont
  {E.}~\bibnamefont {Huffman}}, \bibinfo {author} {\bibfnamefont
  {Z.}~\bibnamefont {Liu}}, \bibinfo {author} {\bibfnamefont {F.}~\bibnamefont
  {Parisen~Toldin}}, \bibinfo {author} {\bibfnamefont {J.~S.~E.}\ \bibnamefont
  {Portela}},\ and\ \bibinfo {author} {\bibfnamefont {J.}~\bibnamefont
  {Schwab}} (\bibinfo {collaboration} {ALF}),\ }\bibfield  {title} {\bibinfo
  {title} {The {ALF} ({Algorithms} for {Lattice} {Fermions}) project release
  2.4. {Documentation} for the auxiliary-field quantum {Monte} {Carlo} code},\
  }\href {https://doi.org/10.21468/SciPostPhysCodeb.1} {\bibfield  {journal}
  {\bibinfo  {journal} {SciPost Phys. Codebases}\ }\textbf {\bibinfo {volume}
  {2022}},\ \bibinfo {pages} {1} (\bibinfo {year} {2022})},\ \Eprint
  {https://arxiv.org/abs/2012.11914} {arXiv:2012.11914 [cond-mat.str-el]}
  \BibitemShut {NoStop}%
\bibitem [{\citenamefont {Zheng}\ \emph {et~al.}(2011)\citenamefont {Zheng},
  \citenamefont {Zhang},\ and\ \citenamefont {Wu}}]{zheng2011particle}%
  \BibitemOpen
  \bibfield  {author} {\bibinfo {author} {\bibfnamefont {D.}~\bibnamefont
  {Zheng}}, \bibinfo {author} {\bibfnamefont {G.-M.}\ \bibnamefont {Zhang}},\
  and\ \bibinfo {author} {\bibfnamefont {C.}~\bibnamefont {Wu}},\ }\bibfield
  {title} {\bibinfo {title} {Particle-hole symmetry and interaction effects in
  the {Kane}-{Mele}-{Hubbard} model},\ }\href
  {https://doi.org/10.1103/PhysRevB.84.205121} {\bibfield  {journal} {\bibinfo
  {journal} {Phys. Rev. B}\ }\textbf {\bibinfo {volume} {84}},\ \bibinfo
  {pages} {205121} (\bibinfo {year} {2011})}\BibitemShut {NoStop}%
\bibitem [{\citenamefont {Li}\ and\ \citenamefont {Yao}(2017)}]{li2017edge}%
  \BibitemOpen
  \bibfield  {author} {\bibinfo {author} {\bibfnamefont {Z.-X.}\ \bibnamefont
  {Li}}\ and\ \bibinfo {author} {\bibfnamefont {H.}~\bibnamefont {Yao}},\
  }\bibfield  {title} {\bibinfo {title} {Edge stability and edge quantum
  criticality in two-dimensional interacting topological insulators},\ }\href
  {https://doi.org/10.1103/PhysRevB.96.241101} {\bibfield  {journal} {\bibinfo
  {journal} {Phys. Rev. B}\ }\textbf {\bibinfo {volume} {96}},\ \bibinfo
  {pages} {241101} (\bibinfo {year} {2017})}\BibitemShut {NoStop}%
\bibitem [{\citenamefont {Hohenadler}\ \emph {et~al.}(2011)\citenamefont
  {Hohenadler}, \citenamefont {Lang},\ and\ \citenamefont
  {Assaad}}]{hohenadler2011correlation}%
  \BibitemOpen
  \bibfield  {author} {\bibinfo {author} {\bibfnamefont {M.}~\bibnamefont
  {Hohenadler}}, \bibinfo {author} {\bibfnamefont {T.~C.}\ \bibnamefont
  {Lang}},\ and\ \bibinfo {author} {\bibfnamefont {F.~F.}\ \bibnamefont
  {Assaad}},\ }\bibfield  {title} {\bibinfo {title} {Correlation effects in
  quantum spin-{Hall} insulators: {A} quantum {Monte} {Carlo} study},\ }\href
  {https://doi.org/10.1103/PhysRevLett.106.100403} {\bibfield  {journal}
  {\bibinfo  {journal} {Phys. Rev. Lett.}\ }\textbf {\bibinfo {volume} {106}},\
  \bibinfo {pages} {100403} (\bibinfo {year} {2011})}\BibitemShut {NoStop}%
\bibitem [{\citenamefont {Hohenadler}\ \emph {et~al.}(2012)\citenamefont
  {Hohenadler}, \citenamefont {Meng}, \citenamefont {Lang}, \citenamefont
  {Wessel}, \citenamefont {Muramatsu},\ and\ \citenamefont
  {Assaad}}]{hohenadler2012qptkmh}%
  \BibitemOpen
  \bibfield  {author} {\bibinfo {author} {\bibfnamefont {M.}~\bibnamefont
  {Hohenadler}}, \bibinfo {author} {\bibfnamefont {Z.~Y.}\ \bibnamefont
  {Meng}}, \bibinfo {author} {\bibfnamefont {T.~C.}\ \bibnamefont {Lang}},
  \bibinfo {author} {\bibfnamefont {S.}~\bibnamefont {Wessel}}, \bibinfo
  {author} {\bibfnamefont {A.}~\bibnamefont {Muramatsu}},\ and\ \bibinfo
  {author} {\bibfnamefont {F.~F.}\ \bibnamefont {Assaad}},\ }\bibfield  {title}
  {\bibinfo {title} {Quantum phase transitions in the {Kane}-{Mele}-{Hubbard}
  model},\ }\href {https://doi.org/10.1103/PhysRevB.85.115132} {\bibfield
  {journal} {\bibinfo  {journal} {Phys. Rev. B}\ }\textbf {\bibinfo {volume}
  {85}},\ \bibinfo {pages} {115132} (\bibinfo {year} {2012})}\BibitemShut
  {NoStop}%
\bibitem [{\citenamefont {Hohenadler}\ and\ \citenamefont
  {Assaad}(2013)}]{hohenadler2013correlation}%
  \BibitemOpen
  \bibfield  {author} {\bibinfo {author} {\bibfnamefont {M.}~\bibnamefont
  {Hohenadler}}\ and\ \bibinfo {author} {\bibfnamefont {F.~F.}\ \bibnamefont
  {Assaad}},\ }\bibfield  {title} {\bibinfo {title} {Correlation effects in
  two-dimensional topological insulators},\ }\href
  {https://doi.org/10.1088/0953-8984/25/14/143201} {\bibfield  {journal}
  {\bibinfo  {journal} {J. Condens. Matter Phys.}\ }\textbf {\bibinfo {volume}
  {25}},\ \bibinfo {pages} {143201} (\bibinfo {year} {2013})}\BibitemShut
  {NoStop}%
\bibitem [{\citenamefont {Rachel}(2018)}]{rachel2018review}%
  \BibitemOpen
  \bibfield  {author} {\bibinfo {author} {\bibfnamefont {S.}~\bibnamefont
  {Rachel}},\ }\bibfield  {title} {\bibinfo {title} {Interacting topological
  insulators: a review},\ }\href {https://doi.org/10.1088/1361-6633/aad6a6}
  {\bibfield  {journal} {\bibinfo  {journal} {Rep. Prog. Phys.}\ }\textbf
  {\bibinfo {volume} {81}},\ \bibinfo {pages} {116501} (\bibinfo {year}
  {2018})}\BibitemShut {NoStop}%
\bibitem [{\citenamefont {Li}\ \emph {et~al.}(2015)\citenamefont {Li},
  \citenamefont {Jiang},\ and\ \citenamefont {Yao}}]{li2015majorana}%
  \BibitemOpen
  \bibfield  {author} {\bibinfo {author} {\bibfnamefont {Z.-X.}\ \bibnamefont
  {Li}}, \bibinfo {author} {\bibfnamefont {Y.-F.}\ \bibnamefont {Jiang}},\ and\
  \bibinfo {author} {\bibfnamefont {H.}~\bibnamefont {Yao}},\ }\bibfield
  {title} {\bibinfo {title} {Solving the fermion sign problem in quantum
  {Monte} {Carlo} simulations by {Majorana} representation},\ }\href
  {https://doi.org/10.1103/PhysRevB.91.241117} {\bibfield  {journal} {\bibinfo
  {journal} {Phys. Rev. B}\ }\textbf {\bibinfo {volume} {91}},\ \bibinfo
  {pages} {241117} (\bibinfo {year} {2015})}\BibitemShut {NoStop}%
\bibitem [{\citenamefont {Li}\ \emph {et~al.}(2016)\citenamefont {Li},
  \citenamefont {Jiang},\ and\ \citenamefont {Yao}}]{li2016majorana}%
  \BibitemOpen
  \bibfield  {author} {\bibinfo {author} {\bibfnamefont {Z.-X.}\ \bibnamefont
  {Li}}, \bibinfo {author} {\bibfnamefont {Y.-F.}\ \bibnamefont {Jiang}},\ and\
  \bibinfo {author} {\bibfnamefont {H.}~\bibnamefont {Yao}},\ }\bibfield
  {title} {\bibinfo {title} {Majorana-time-reversal symmetries: A fundamental
  principle for sign-problem-free quantum {Monte} {Carlo} simulations},\ }\href
  {https://doi.org/10.1103/PhysRevLett.117.267002} {\bibfield  {journal}
  {\bibinfo  {journal} {Phys. Rev. Lett.}\ }\textbf {\bibinfo {volume} {117}},\
  \bibinfo {pages} {267002} (\bibinfo {year} {2016})}\BibitemShut {NoStop}%
\bibitem [{\citenamefont {Hohenadler}\ and\ \citenamefont
  {Assaad}(2014)}]{hohenadler2014rashba}%
  \BibitemOpen
  \bibfield  {author} {\bibinfo {author} {\bibfnamefont {M.}~\bibnamefont
  {Hohenadler}}\ and\ \bibinfo {author} {\bibfnamefont {F.~F.}\ \bibnamefont
  {Assaad}},\ }\bibfield  {title} {\bibinfo {title} {Rashba coupling and
  magnetic order in correlated helical liquids},\ }\href
  {https://doi.org/10.1103/PhysRevB.90.245148} {\bibfield  {journal} {\bibinfo
  {journal} {Phys. Rev. B}\ }\textbf {\bibinfo {volume} {90}},\ \bibinfo
  {pages} {245148} (\bibinfo {year} {2014})}\BibitemShut {NoStop}%
\bibitem [{\citenamefont {Bychkov}\ and\ \citenamefont
  {Rashba}(1984)}]{bychkov1984oscillatory}%
  \BibitemOpen
  \bibfield  {author} {\bibinfo {author} {\bibfnamefont {Y.~A.}\ \bibnamefont
  {Bychkov}}\ and\ \bibinfo {author} {\bibfnamefont {E.~I.}\ \bibnamefont
  {Rashba}},\ }\bibfield  {title} {\bibinfo {title} {Oscillatory effects and
  the magnetic susceptibility of carriers in inversion layers},\ }\href
  {https://doi.org/10.1088/0022-3719/17/33/015} {\bibfield  {journal} {\bibinfo
   {journal} {J. Phys. C: Solid State Phys.}\ }\textbf {\bibinfo {volume}
  {17}},\ \bibinfo {pages} {6039} (\bibinfo {year} {1984})}\BibitemShut
  {NoStop}%
\bibitem [{Note1()}]{Note1}%
  \BibitemOpen
  \bibinfo {note} {The ground state is degenerate and thus naively precludes
  the use of the projective DQMC. See the Supplemental~\cite
  {supplemental}.}\BibitemShut {Stop}%
\bibitem [{\citenamefont {Melchert}(2009)}]{autoscale}%
  \BibitemOpen
  \bibfield  {author} {\bibinfo {author} {\bibfnamefont {O.}~\bibnamefont
  {Melchert}},\ }\bibfield  {title} {\bibinfo {title} {{autoScale.py} --- a
  program for automatic finite-size scaling analyses: A user's guide},\
  }\Eprint {https://arxiv.org/abs/0910.5403} {arXiv:0910.5403
  [physics.comp-ph]}  (\bibinfo {year} {2009})\BibitemShut {NoStop}%
\bibitem [{\citenamefont {Houdayer}\ and\ \citenamefont
  {Hartmann}(2004)}]{houdayer2004collapse}%
  \BibitemOpen
  \bibfield  {author} {\bibinfo {author} {\bibfnamefont {J.}~\bibnamefont
  {Houdayer}}\ and\ \bibinfo {author} {\bibfnamefont {A.~K.}\ \bibnamefont
  {Hartmann}},\ }\bibfield  {title} {\bibinfo {title} {Low-temperature behavior
  of two-dimensional {Gaussian} {Ising} spin glasses},\ }\href
  {https://doi.org/10.1103/PhysRevB.70.014418} {\bibfield  {journal} {\bibinfo
  {journal} {Phys. Rev. B}\ }\textbf {\bibinfo {volume} {70}},\ \bibinfo
  {pages} {014418} (\bibinfo {year} {2004})}\BibitemShut {NoStop}%
\bibitem [{sup()}]{supplemental}%
  \BibitemOpen
  \href@noop {} {}\bibinfo {note} {See the Supplemental Material for details on
  the determinant quantum Monte Carlo simulations, and the helical edge
  state.}\BibitemShut {Stop}%
\bibitem [{Note2()}]{Note2}%
  \BibitemOpen
  \bibinfo {note} {Our labels of $g_i$ couplings are different from the other
  bosonization conventions.}\BibitemShut {Stop}%
\bibitem [{\citenamefont {Doh}\ and\ \citenamefont
  {Jeon}(2013)}]{doh2013bifurcation}%
  \BibitemOpen
  \bibfield  {author} {\bibinfo {author} {\bibfnamefont {H.}~\bibnamefont
  {Doh}}\ and\ \bibinfo {author} {\bibfnamefont {G.~S.}\ \bibnamefont {Jeon}},\
  }\bibfield  {title} {\bibinfo {title} {Bifurcation of the edge-state width in
  a two-dimensional topological insulator},\ }\href
  {https://doi.org/10.1103/PhysRevB.88.245115} {\bibfield  {journal} {\bibinfo
  {journal} {Phys. Rev. B}\ }\textbf {\bibinfo {volume} {88}},\ \bibinfo
  {pages} {245115} (\bibinfo {year} {2013})}\BibitemShut {NoStop}%
\bibitem [{Note3()}]{Note3}%
  \BibitemOpen
  \bibinfo {note} {More precisely, at the ordinary transition, the boundary
  primary field is the normal derivative of the order parameter field. Here, we
  denote it as $\protect \hat \phi $ for simplicity.}\BibitemShut {Stop}%
\bibitem [{\citenamefont {Deng}\ and\ \citenamefont
  {Bl\"ote}(2003)}]{deng2003bulk}%
  \BibitemOpen
  \bibfield  {author} {\bibinfo {author} {\bibfnamefont {Y.}~\bibnamefont
  {Deng}}\ and\ \bibinfo {author} {\bibfnamefont {H.~W.~J.}\ \bibnamefont
  {Bl\"ote}},\ }\bibfield  {title} {\bibinfo {title} {Bulk and surface critical
  behavior of the three-dimensional {Ising} model and conformal invariance},\
  }\href {https://doi.org/10.1103/PhysRevE.67.066116} {\bibfield  {journal}
  {\bibinfo  {journal} {Phys. Rev. E}\ }\textbf {\bibinfo {volume} {67}},\
  \bibinfo {pages} {066116} (\bibinfo {year} {2003})}\BibitemShut {NoStop}%
\bibitem [{\citenamefont {Deng}\ \emph {et~al.}(2005)\citenamefont {Deng},
  \citenamefont {Bl\"ote},\ and\ \citenamefont
  {Nightingale}}]{deng2005surface}%
  \BibitemOpen
  \bibfield  {author} {\bibinfo {author} {\bibfnamefont {Y.}~\bibnamefont
  {Deng}}, \bibinfo {author} {\bibfnamefont {H.~W.~J.}\ \bibnamefont
  {Bl\"ote}},\ and\ \bibinfo {author} {\bibfnamefont {M.~P.}\ \bibnamefont
  {Nightingale}},\ }\bibfield  {title} {\bibinfo {title} {Surface and bulk
  transitions in three-dimensional $\mathrm{O}(n)$ models},\ }\href
  {https://doi.org/10.1103/PhysRevE.72.016128} {\bibfield  {journal} {\bibinfo
  {journal} {Phys. Rev. E}\ }\textbf {\bibinfo {volume} {72}},\ \bibinfo
  {pages} {016128} (\bibinfo {year} {2005})}\BibitemShut {NoStop}%
\bibitem [{\citenamefont {Gliozzi}\ \emph {et~al.}(2015)\citenamefont
  {Gliozzi}, \citenamefont {Liendo}, \citenamefont {Meineri},\ and\
  \citenamefont {Rago}}]{gliozzi2015boundary}%
  \BibitemOpen
  \bibfield  {author} {\bibinfo {author} {\bibfnamefont {F.}~\bibnamefont
  {Gliozzi}}, \bibinfo {author} {\bibfnamefont {P.}~\bibnamefont {Liendo}},
  \bibinfo {author} {\bibfnamefont {M.}~\bibnamefont {Meineri}},\ and\ \bibinfo
  {author} {\bibfnamefont {A.}~\bibnamefont {Rago}},\ }\bibfield  {title}
  {\bibinfo {title} {Boundary and interface {CFTs} from the conformal
  bootstrap},\ }\href {https://doi.org/10.1007/JHEP05(2015)036} {\bibfield
  {journal} {\bibinfo  {journal} {JHEP}\ }\textbf {\bibinfo {volume} {05}},\
  \bibinfo {pages} {036}},\ \bibinfo {note} {[Erratum:
  \href{https://doi.org/10.1007/JHEP12(2021)093}{JHEP 12, 093 (2021)}]},\
  \Eprint {https://arxiv.org/abs/1502.07217} {arXiv:1502.07217 [hep-th]}
  \BibitemShut {NoStop}%
\bibitem [{\citenamefont {Zhou}\ and\ \citenamefont
  {Zou}(2025)}]{zhou2024studying}%
  \BibitemOpen
  \bibfield  {author} {\bibinfo {author} {\bibfnamefont {Z.}~\bibnamefont
  {Zhou}}\ and\ \bibinfo {author} {\bibfnamefont {Y.}~\bibnamefont {Zou}},\
  }\bibfield  {title} {\bibinfo {title} {{Studying the 3d {Ising} surface
  {CFTs} on the fuzzy sphere}},\ }\href
  {https://doi.org/10.21468/SciPostPhys.18.1.031} {\bibfield  {journal}
  {\bibinfo  {journal} {SciPost Phys.}\ }\textbf {\bibinfo {volume} {18}},\
  \bibinfo {pages} {031} (\bibinfo {year} {2025})},\ \Eprint
  {https://arxiv.org/abs/2407.15914} {arXiv:2407.15914 [hep-th]} \BibitemShut
  {NoStop}%
\bibitem [{\citenamefont {Dedushenko}(2024)}]{dedushenko2024ising}%
  \BibitemOpen
  \bibfield  {author} {\bibinfo {author} {\bibfnamefont {M.}~\bibnamefont
  {Dedushenko}},\ }\bibfield  {title} {\bibinfo {title} {Ising {BCFT} from
  fuzzy hemisphere},\ }\Eprint {https://arxiv.org/abs/2407.15948}
  {arXiv:2407.15948 [hep-th]}  (\bibinfo {year} {2024})\BibitemShut {NoStop}%
\bibitem [{Note4()}]{Note4}%
  \BibitemOpen
  \bibinfo {note} {Note that at this point, the two-particle scattering process
  is still irrelevant.}\BibitemShut {Stop}%
\bibitem [{Note5()}]{Note5}%
  \BibitemOpen
  \bibinfo {note} {This prediction is not accurate given the renormalization
  effect from the strong fluctuations in the bulk. We will see the effect in
  the nonperturbative Monte Carlo calculations.}\BibitemShut {Stop}%
\bibitem [{\citenamefont {Grover}\ \emph {et~al.}(2014)\citenamefont {Grover},
  \citenamefont {Sheng},\ and\ \citenamefont
  {Vishwanath}}]{grover2013emergent}%
  \BibitemOpen
  \bibfield  {author} {\bibinfo {author} {\bibfnamefont {T.}~\bibnamefont
  {Grover}}, \bibinfo {author} {\bibfnamefont {D.~N.}\ \bibnamefont {Sheng}},\
  and\ \bibinfo {author} {\bibfnamefont {A.}~\bibnamefont {Vishwanath}},\
  }\bibfield  {title} {\bibinfo {title} {{Emergent Space-Time Supersymmetry at
  the Boundary of a Topological Phase}},\ }\href
  {https://doi.org/10.1126/science.1248253} {\bibfield  {journal} {\bibinfo
  {journal} {Science}\ }\textbf {\bibinfo {volume} {344}},\ \bibinfo {pages}
  {280} (\bibinfo {year} {2014})},\ \Eprint {https://arxiv.org/abs/1301.7449}
  {arXiv:1301.7449 [cond-mat.str-el]} \BibitemShut {NoStop}%
\bibitem [{\citenamefont {Li}\ \emph {et~al.}(2017)\citenamefont {Li},
  \citenamefont {Jiang},\ and\ \citenamefont {Yao}}]{li2017edge2}%
  \BibitemOpen
  \bibfield  {author} {\bibinfo {author} {\bibfnamefont {Z.-X.}\ \bibnamefont
  {Li}}, \bibinfo {author} {\bibfnamefont {Y.-F.}\ \bibnamefont {Jiang}},\ and\
  \bibinfo {author} {\bibfnamefont {H.}~\bibnamefont {Yao}},\ }\bibfield
  {title} {\bibinfo {title} {Edge quantum criticality and emergent
  supersymmetry in topological phases},\ }\href
  {https://doi.org/10.1103/PhysRevLett.119.107202} {\bibfield  {journal}
  {\bibinfo  {journal} {Phys. Rev. Lett.}\ }\textbf {\bibinfo {volume} {119}},\
  \bibinfo {pages} {107202} (\bibinfo {year} {2017})}\BibitemShut {NoStop}%
\bibitem [{\citenamefont {Jia}\ \emph {et~al.}(2022)\citenamefont {Jia},
  \citenamefont {Marcellina}, \citenamefont {Das}, \citenamefont {Lodge},
  \citenamefont {Wang}, \citenamefont {Ho}, \citenamefont {Biswas},
  \citenamefont {Pham}, \citenamefont {Tao}, \citenamefont {Huang},
  \citenamefont {Lin}, \citenamefont {Bansil}, \citenamefont {Mukherjee},\ and\
  \citenamefont {Weber}}]{jia2022tuning}%
  \BibitemOpen
  \bibfield  {author} {\bibinfo {author} {\bibfnamefont {J.}~\bibnamefont
  {Jia}}, \bibinfo {author} {\bibfnamefont {E.}~\bibnamefont {Marcellina}},
  \bibinfo {author} {\bibfnamefont {A.}~\bibnamefont {Das}}, \bibinfo {author}
  {\bibfnamefont {M.~S.}\ \bibnamefont {Lodge}}, \bibinfo {author}
  {\bibfnamefont {B.}~\bibnamefont {Wang}}, \bibinfo {author} {\bibfnamefont
  {D.-Q.}\ \bibnamefont {Ho}}, \bibinfo {author} {\bibfnamefont
  {R.}~\bibnamefont {Biswas}}, \bibinfo {author} {\bibfnamefont {T.~A.}\
  \bibnamefont {Pham}}, \bibinfo {author} {\bibfnamefont {W.}~\bibnamefont
  {Tao}}, \bibinfo {author} {\bibfnamefont {C.-Y.}\ \bibnamefont {Huang}},
  \bibinfo {author} {\bibfnamefont {H.}~\bibnamefont {Lin}}, \bibinfo {author}
  {\bibfnamefont {A.}~\bibnamefont {Bansil}}, \bibinfo {author} {\bibfnamefont
  {S.}~\bibnamefont {Mukherjee}},\ and\ \bibinfo {author} {\bibfnamefont
  {B.}~\bibnamefont {Weber}},\ }\bibfield  {title} {\bibinfo {title} {Tuning
  the many-body interactions in a helical {Luttinger} liquid},\ }\href
  {https://doi.org/10.1038/s41467-022-33676-0} {\bibfield  {journal} {\bibinfo
  {journal} {Nat. Commun.}\ }\textbf {\bibinfo {volume} {13}},\ \bibinfo
  {pages} {6046} (\bibinfo {year} {2022})}\BibitemShut {NoStop}%
\end{thebibliography}%

	\appendix
	\onecolumngrid
	
	\setcounter{secnumdepth}{3}
	\setcounter{equation}{0}
	\setcounter{figure}{0}
	\renewcommand{\theequation}{S\arabic{equation}}
	\renewcommand{\thefigure}{S\arabic{figure}}
	\renewcommand\figurename{Supplementary Figure}
	\renewcommand\tablename{Supplementary Table}

	\section{Details of the determinant quantum Monte Carlo simulation}
	
	In the determinant quantum Monte Carlo formulation of purely fermionic problems, each $\boldsymbol{i}$th local interaction term is decoupled into fermion bilinears using real Hubbard--Stratonivich fields $\phi_{\boldsymbol{i}}(\tau)$ \cite{assaad2020ALF}. This results in an action bilinear in the fermions, which is then integrated into a determinant that depends on $\{\phi_{\boldsymbol{i}}(\tau)\}$. Monte Carlo integration over $\phi_{\boldsymbol{i}}(\tau)$ then samples the desired Green's function and observables. In our calculation, the resultant Gaussian integral over $\phi_{\boldsymbol{i}}(\tau)$ is replaced by the fourth order Gauss–Hermite quadrature \cite{assaad2020ALF}, summed at roots of the Hermite polynomial $\{\eta(l)\}$ with weights $\{\gamma(l)\}$.
	
	To avoid the sign problem, i.e., negative weights in the Monte Carlo sampling, we decouple the interactions so that the Hamiltonian stays invariant under the following antiunitary symmetry,
	\begin{equation}
		\label{eq:ss:Theta}
		\Theta: c_{\boldsymbol{j} \sigma} \rightarrow (- 1)^{s(\boldsymbol{j})} c_{\boldsymbol{j}
			\bar{\sigma}}^{\dag},
		\quad c_{\boldsymbol{j} \sigma}^{\dag} \rightarrow (- 1)^{s(\boldsymbol{j})} c_{\boldsymbol{j} \bar{\sigma}},
		\quad i \to - i,
	\end{equation}
	where $s(\boldsymbol{j})=0,1$ on the two sublattice sites, $\sigma=\uparrow, \downarrow$ denotes the spin, and $\bar{\sigma}$ denotes the opposite spin. The Hubbard and Rashba interaction terms in Eq.~\myeqref{eq:Hint} are decoupled via
	\begin{gather}
		\exp \left[ -U\Delta \tau \left( c_{\boldsymbol{i}\uparrow}^\dag c_{\boldsymbol{i}\uparrow} - \frac12\right) \left( c_{\boldsymbol{i}\downarrow}^\dag c_{\boldsymbol{i}\downarrow} - \frac12\right) - \frac{U\Delta\tau}{4} \right] %
		=  \frac{1}{4}  \sum_l \gamma_{\boldsymbol i} (l) \exp \left[ i \sqrt{\frac{U \Delta \tau}{2}}
		\eta_{\boldsymbol i} (l) (c_{\boldsymbol i \uparrow}^{\dag} c_{\boldsymbol i \uparrow} + c_{\boldsymbol i \downarrow}^{\dag}
		c_{\boldsymbol i \downarrow} - 1) \right], \\ %
		\exp [- V \Delta \tau (c_{\boldsymbol{i}\uparrow}^{\dag} c_{\boldsymbol{j}\uparrow}^{\dag}
		c_{\boldsymbol{i}\downarrow} c_{\boldsymbol{j}\downarrow} + \text{H.c.})]
		=  \frac{1}{4}  \sum_{l,\pm} \gamma^R_{\boldsymbol{i}\boldsymbol{j}, \pm} (l) 
		\exp \left[ i \sqrt{\frac{V \Delta \tau}{4}} \eta_{\boldsymbol i \boldsymbol j, \pm} (l) (c_{\boldsymbol i \uparrow}^{\dag} c_{\boldsymbol j \uparrow}^{\dag} \pm c_{\boldsymbol j \uparrow} c_{\boldsymbol i \uparrow} +
		c_{\boldsymbol i \downarrow} c_{\boldsymbol j \downarrow} \pm c_{\boldsymbol j \downarrow}^{\dag} c_{\boldsymbol i \downarrow}^{\dag}) \right]. 
	\end{gather}
	Note that $U, V\ge0$. Even though both decouplings appear complex, they are sign-problem-free since the weights contributed by the two spin sectors are complex conjugates of each other due to $\Theta$ in \eqref{eq:ss:Theta}. In general, decoupling in the complex channel, e.g.\ for the Hubbard interaction, improves the condition number of the linear algebra at low temperature.
	
	To implement the decouplng scheme above, we introduce the Nambu spinors to accommodate the pairing terms due to the Rashba interaction, defined by $\Psi_{\boldsymbol i \sigma} \equiv ( c_{\boldsymbol i \sigma}, c_{\boldsymbol i \sigma}^{\dag})^T$. Denote by $\sigma$ and $\mu$ the Pauli matrices acting in spin and particle-hole space, respectively, the spin operator in Nambu basis reads
	\begin{equation}
		\vec S_{\boldsymbol i} = \frac14 \Psi^\dag_{\boldsymbol i} \left( \mu^z \sigma^x, \sigma^y, \mu^z \sigma^z \right) \Psi_{\boldsymbol i},
	\end{equation}
	where the spin indices of $\Psi$ and $\sigma$ are summed over and suppressed. The factor $1/4$ is due to the spin $1/2$ and the double counting of Nambu components. After the decoupling, the sampled action is quadratic in fermion fields, and Wick contraction is used to evaluate spin-spin correlators. We define the equal-time correlation functions
	\begin{equation}
		G^>_{\boldsymbol i\sigma,\boldsymbol j\sigma'} = \langle \Psi_{\boldsymbol i\sigma} \Psi^\dag_{\boldsymbol j\sigma'} \rangle, \quad G^<_{\boldsymbol i\sigma,\boldsymbol j\sigma'} = \langle \Psi^\dag_{\boldsymbol i \sigma} \Psi_{\boldsymbol j\sigma'} \rangle.
	\end{equation}
	Denote the spin-up Green's functions by
	\begin{equation}
		G^>_{\boldsymbol i \boldsymbol j} \equiv \langle \Psi_{\boldsymbol i\uparrow} \Psi^\dag_{\boldsymbol j\uparrow} \rangle, \quad G^<_{\boldsymbol i \boldsymbol j} \equiv \langle \Psi^\dag_{\boldsymbol i\uparrow} \Psi_{\boldsymbol j\uparrow} \rangle.
	\end{equation}
	Then, according to the symmetry transformation $\Theta$,
	\begin{equation}
		G^>_{\boldsymbol i\downarrow,\boldsymbol j\downarrow} = (-1)^{s(\boldsymbol i)+ s(\boldsymbol j)} (G^<_{\boldsymbol i \boldsymbol j})^\ast, \quad G^<_{\boldsymbol i\downarrow, \boldsymbol j\downarrow} = (-1)^{s(\boldsymbol i)+ s(\boldsymbol j)} (G^>_{\boldsymbol i \boldsymbol j})^\ast.
	\end{equation}
	Note that these equal-spin matrices are indexed by the Nambu components. Unequal-spin Green's functions are zero for the Monte Carlo sampling.
	The spin-spin correlator is given by
	\begin{equation}
		\langle S^y_{\boldsymbol i} S^y_{\boldsymbol j} \rangle = \frac{(-1)^{s(\boldsymbol i)+ s(\boldsymbol j)}}{16} \left(  \Tr \left[  (G^>_{\boldsymbol i \boldsymbol j})^\ast  (G^>_{\boldsymbol i \boldsymbol j})^T \right] + \Tr \left[  G^<_{\boldsymbol i \boldsymbol j}  (G^<_{\boldsymbol i \boldsymbol j})^\dag \right] +\Tr \left[  (G^<_{\boldsymbol i \boldsymbol j})^\ast  (G^<_{\boldsymbol i \boldsymbol j})^T \right] + \Tr \left[ G^>_{\boldsymbol i \boldsymbol j} (G^>_{\boldsymbol i \boldsymbol j})^\dag \right] \right).
	\end{equation}
	
	Using the Nambu spinors double the number of fermions. On the other hand, the two spin sectors are related by $\Theta$. Therefore, the total action need only be evaluated with the Nambu spinors in the spin-up sector, to account for the correct sampling weight.
	
	For the simulations in this paper, we set $t=1$, $\lambda = 0.3$ and $V=0.1$. We simulated at a very low temperature, namely $\beta=120$ for all system sizes, except for $L=12$ where $\beta=85$ is used. We used the finite-temperature DQMC instead of the projective version because the ground state is degenerate in the disordered phase due to the symmetry $c_{\boldsymbol i \sigma} \to (-1)^{s(\boldsymbol{i})}c^\dag_{\boldsymbol i \sigma}$. Consequently, the projective DQMC does not naively fit this system, although it produces critical fields and scaling exponents similar to our result. In particular, for even $L$'s, the symmetry enforces $G_\text{bdy}(L/2) = 0$ at the boundary as shown in Fig.~\ref*{fig:special}(b). However, this is typically violated by the projective DQMC naively.
	
	Finally, we outline the procedure for the scaling analysis. For the RG-invariant quantity $R^{SS}_\mathrm{w}$ vs.\ interaction strength curves at different system sizes in Fig.~\ref*{fig:crossing}(a)--(d), we used a modified version of the \textsc{autoScale.py} package~\cite{autoscale} to obtain the critical field and scaling exponents. In brief, for each scaled data point $p:(\tilde{x},\tilde{y})$, the method finds the two closest data points in each other system sizes that brackets $p$. The deviation of $p$ from these bracketing line segments of different system sizes gives a $\chi^2$ error. Minimizing the total $\chi^2$ leads to estimates for the scaling exponents and correlation lengths. We extended \textsc{autoScale.py} to handle logarithmic scalings.
	
	We note that much stronger finite-size effects appear in the boundary transitions for $U_\text{bulk}\gtrsim U^*_\text{bulk}$. As a consequence, $U_\text{bdy}$ at the crossing shifts visibly and monotonically as the system size increases, as shown in the insets of Fig.~\ref*{fig:crossing}(e) and (f). A linear extrapolation of the $U_\text{bdy}$ of the crossing with inverse system size $1/L\to0$ is used to extract the interaction amplitude at the transition, with a negative value indicating that the boundary orders at $U_\text{bdy}=0$.
	
	A different method is needed to extract the Luttinger parameter in Figs.~\ref*{fig:crossing} and \ref*{fig:special} from the spatial correlators, since there is no crossing.
	For the correlators, we scale the data points into $q:\left[\sin(\tfrac{\pi x}{L}), y L^{2\Delta}\right]$. At each system size $L$, the points at $x=0,\pm1,\pm2$ are dropped since they are dominated by UV physics. Then, a fourth-order polynomial is fitted through the scaled data points of all system sizes. Minimizing the total $\chi^2$ with respect to $\Delta$ again produces our estimates. An $S+1$ analysis~\cite{houdayer2004collapse} determines the uncertainty, similar to the method in \textsc{autoScale.py}.
	
	\section{Analytical derivation for the Luttinger parameter of the helical edge state}
	
\begin{figure}
		\centering
		\subfigure[]{\includegraphics[width=0.4\linewidth]{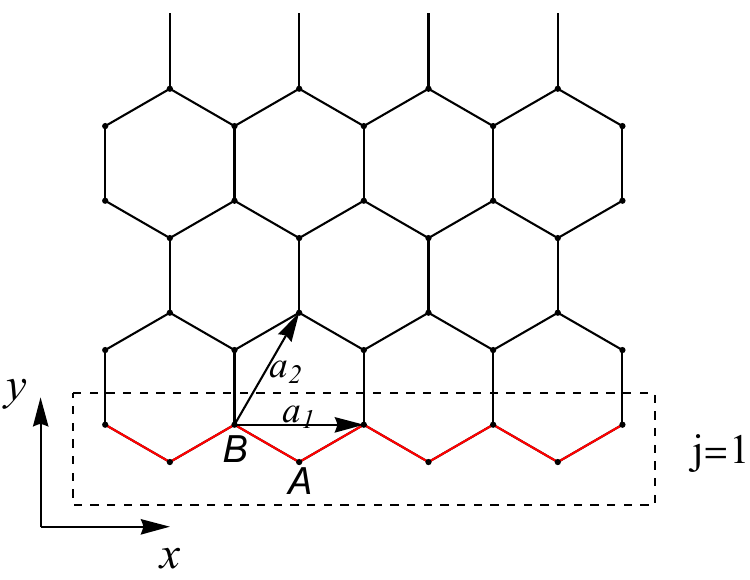}} \quad
		\subfigure[]
		{\includegraphics[width=0.4\linewidth]{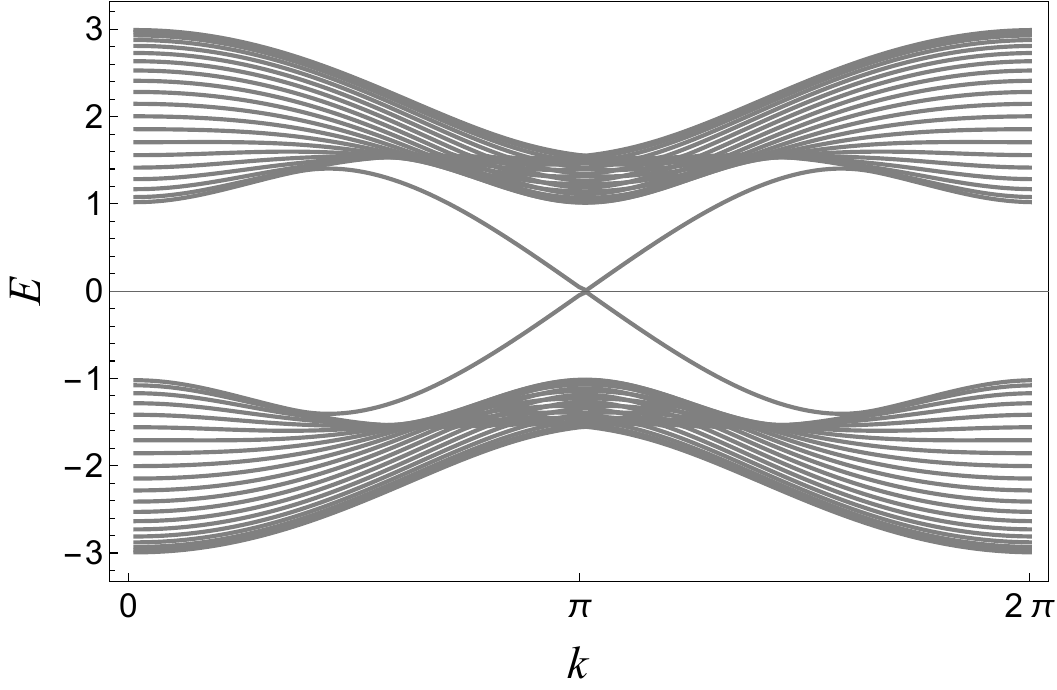}}
		\caption{(a) Illustration of the lattice with a zigzag boundary. $A$ and $B$ denote the two sublattices, respectively. 
			$a_{1,2}$ is the unit vector for the honeycomb lattice. 
			(b) Energy-momentum spectrum of the noninteracting Kane-Mele model showing gapped bulk bands and gapless edge modes. Helical edge modes of two spins cross zero energy at $k=\pi$.}
		\label{fig:lattice_sm} 
	\end{figure}

	We consider the Kane-Mele model with zigzag boundaries, as shown in Fig.~\ref{fig:lattice_sm}(a). 
	The lattice terminates at two zigzag boundaries in
	the $y$ direction and is periodic in the $x$ direction. 
	Note that only one zigzag boundary is shown in Fig.~\ref{fig:lattice_sm}(a).  
	Due to the nontrivial topology, the Kane-Mele model features topological edge states. 
	We plot the spectrum of the Kane-Mele model in Fig.~\ref{fig:lattice_sm}(b), with $\lambda = 0.3$ and $t=1$. 
	As expected, there are helical edge states that intersect at momentum $k_x = \pi$. 
	Next, we will derive the effective theory for the helical edge states near momentum $\pi$.
	Our derivation closely follows Ref.~\onlinecite{doh2013bifurcation}. 
	
	To analytically derive the gapless helical edge state, we consider a semi-infinite honeycomb lattice as shown in Fig.~\ref{fig:lattice_sm}(a). 
	The lattice is periodic in the $x$ direction and hosts a zigzag boundary at $y=0$. 
	We take the two primitive vectors to be $\vec a_1 = (1,0)$ and $\vec a_2 = (\frac12 , \frac{\sqrt3}2)$, in which the lattice constant is set to one for convenience. 
	The Hamiltonian of the noninteracting Kane-Mele model in the basis of $(\psi_{A,\uparrow}, \psi_{B,\uparrow})^T$ is
	\begin{eqnarray}
		H = \left( \ba{cccc} \lambda (-i e^{i k_x} + i e^{i \frac{k_x}2 - i \frac{\sqrt3}2 k_y} + i e^{i \frac{k_x}2 + i \frac{\sqrt{3}}2 k_y } + \text{H.c.}) & - t (1 + e^{i k_x} + e^{i \frac{k_x}2 - i \frac{\sqrt 3}2 k_y} ) \\ - t (1 + e^{-i k_x} + e^{-i \frac{k_x}2 + i \frac{\sqrt 3}2 k_y} ) & -\lambda (-i e^{i k_x} + i e^{i \frac{k_x}2 - i \frac{\sqrt3}2 k_y} + i e^{i \frac{k_x}2 + i \frac{\sqrt{3}}2 k_y}  + \text{H.c.}) 
		\ea \right),
	\end{eqnarray}
	where the momentum $k_x$ and $k_y$ should be understood as quantum operators. 
	Because the spin-up and spin-down components are decoupled in the noninteracting Kane-Mele model, we will first focus on the spin-up component and then obtain the spin-down component via time reversal. 
	The presence of the zigzag boundary breaks the translation symmetry in the $y$ direction, while the translation symmetry in the $x$ direction remains intact. 
	Hence, to get the boundary state, we can consider the eigenstate of the $k_x$ operator and denote it as $k$ for simplicity. 
	The wavefunction ansatz of the localized edge state takes the form, 
	\begin{eqnarray}
		\psi_{j,k} = \left( e^{i \frac{k}2} \Lambda \right)^j \psi_k, 
	\end{eqnarray}
	where $j$ is an integer related to the $y$ coordinate by $y = \frac{\sqrt3}2 j$,  and $\psi_k$ is a two-component wavefunction. 
	The lattice site for the zigzag boundary labeled $j=1$ is illustrated in the box dashed in Fig.~\ref{fig:lattice_sm}(a).
	The factor $\Lambda$ is a constant whose norm is less than one to ensure the convergence of the wavefunction. 
	Note that $\Lambda$ determines the localization length of the edge state. 
	
	With this ansatz, the effective Hamiltonian is 
	\begin{eqnarray}
		H_{\text{eff}} = \left( \ba{cccc} 2 \lambda \sin \frac{k}2 \left( 2 \cos \frac{k}2 - e^{i \frac{k}2} \Lambda - e^{- i \frac{k}2} \Lambda^{-1} \right) & - t \left( 1 + e^{i k} + \Lambda^{-1} \right) \\
		-t \left(1 + e^{-i k} + \Lambda \right) & -2 \lambda \sin \frac{k}2 \left( 2 \cos \frac{k}2 - e^{i \frac{k}2} \Lambda - e^{- i \frac{k}2} \Lambda^{-1} \right) \ea \right).
	\end{eqnarray}
	The dispersion of the boundary wave function and the decay length have been obtained in~\cite{doh2013bifurcation}. 
	Here, we are instead interested in the effective gapless helical edge states near momentum $k=\pi$. 
	Setting $k=\pi$ in the effective Hamiltonian, it is easy to obtain two degenerate zero-energy states, denoted by $\psi^{\pm}_{j,k=\pi}$ with their corresponding decay length given by
	\begin{eqnarray}
		\Lambda_\pm = \pm \frac{\sqrt{t^2 + 16 \lambda^2} - t}{4\lambda}. 
	\end{eqnarray}
	The boundary condition at the zigzag boundary further requires with $b_\pm$ the coefficients of two wavefunctions,
	\begin{eqnarray}
		b_+ \psi^+_{j=0, k=\pi} + b_- \psi^-_{j=0, k=\pi} = 0,
	\end{eqnarray}
	from which we can determine $b_\pm$ and obtain a non-degenerate normalizable zero-energy state located near the zigzag boundary (there is another zero energy state for the spin-down component), 
	\begin{eqnarray} \label{eq:ss:zero_wf}
		\psi_{j,k=\pi} = \left( \frac{2}{\sqrt{1+(4\lambda/t)^2}-1} \right)^{1/2} \sin (j\pi/2) \Lambda^j \left(\ba{cccc} 1 \\ i \Lambda \ea \right), \quad \Lambda = \frac{\sqrt{t^2 + 16 \lambda^2} - t}{4\lambda}. 
	\end{eqnarray}
	
	To obtain an analytical expression for the helical edge states, we carry out a perturbative expansion in momentum. 
	The effective Hamiltonian can be separated into two parts, $H_\text{eff} = H_\text{eff}(k=\pi) + H_\text{eff}(k=\pi + p) $, where the first term is the nonperturbed Hamiltonian and the second term is  the perturbation, $p \ll 1$.
	We treat $\psi_{j,k=\pi}$ as the zero-th order wavefunction, $\psi^{(0)}_j \equiv \psi_{j,k=\pi}$, then the first-order energy is
	\begin{eqnarray}
		E_p &=&  \sum_{j=1}^\infty \Big[ \psi^{(0)\dag}_j \left( \ba{cccc} -2 \lambda p  & i t p\\ -i t p & 2 \lambda p \ea \right) \psi^{(0)}_j +  \Lambda
		\psi^{(0)\dag}_j \left( \ba{cccc} -2 i \lambda + \lambda p  &  0  \\ - t & 2 i \lambda -\lambda p  \ea \right) \psi^{(0)}_j  
		\nn \\
		&& + \Lambda^{-1} \psi^{(0)\dag}_j \left( \ba{cccc} 2 i \lambda + \lambda p  & - t \\ 0 & - 2 i \lambda - \lambda p  \ea \right) \psi^{(0)}_j  \Big]= -v p, \\
		v &=&  \frac{6 \lambda t}{\sqrt{t^2 +(4\lambda)^2}}.
	\end{eqnarray}
	Hence, the spin-up component is a left mover on the bottom zigzag boundary. 
	The spin-down component can be obtained by a time reversal transformation. 
	It turns out that the zero-order wavefunction~\eqref{eq:ss:zero_wf} for the spin-down component remains the same, but the energy changes sign. 
	This indicates that the spin-down component is a right mover, fully consistent with the helical edge state. 
	Projecting to the boundary state, we can express the fermion operator as $c_{\boldsymbol{i}, \sigma} = \psi_{i_y}^{(0)} d_{i_x, \sigma}  $, $\boldsymbol{i} = (i_x, i_y)$. 
	Then the kinetic Hamiltonian in the second-quantized form is 
	\begin{eqnarray}
		\hat H_0 =  \sum_p v p \left(- d^\dag_{p,\uparrow} d_{p,\uparrow} + d^\dag_{p,\downarrow} d_{p,\downarrow} \right), 
	\end{eqnarray}
	where the one-dimensional Fourier transformation is defined by $d_p = \frac1{\sqrt N} d_{i_x} e^{i p i_x}$, where $N$ denotes the number of unit cells along the $x$ axis. 
	The Hubbard interaction becomes
	\begin{eqnarray}
		\hat H_U &=& U_\text{bulk} \sum_{\boldsymbol{i} \in \text{bulk}} \left( c_{\boldsymbol{i},\uparrow}^\dag c_{\boldsymbol{i},\uparrow} - \frac12\right) \left( c_{\boldsymbol{i},\downarrow}^\dag c_{\boldsymbol{i},\downarrow} - \frac12\right)   + U_{\text{bdy}} \sum_{\boldsymbol{i} \in \text{bdy}} \left( c_{\boldsymbol{i},\uparrow}^\dag c_{\boldsymbol{i},\uparrow} - \frac12\right) \left( c_{\boldsymbol{i},\downarrow}^\dag c_{\boldsymbol{i},\downarrow} - \frac12\right) \\
		&=& \frac{U_\text{eff}}{N}  \sum_{p, k_1, k_2} d_{k_1, \uparrow}^\dag  d_{k_1+p, \uparrow}   d_{k_2, \downarrow}^\dag  d_{k_2-p, \downarrow}\,
	\end{eqnarray}
	Here, the effective interaction strength projected onto the helical edge state is given by
	\begin{eqnarray}
		U_\text{eff} &=& U_\text{bulk} \sum_{j=2}^\infty  \left( |\psi_{j,A}^{(0)} |^4 + |\psi_{j,B}^{(0)} |^4 \right) + U_\text{bdy} \left( |\psi_{j=1,A}^{(0)} |^4 + |\psi_{j=1,B}^{(0)} |^4 \right) \\
		&=&  \frac{t}{\sqrt{t^2+(4\lambda)^2}} U_\text{bulk} + \frac{t^2(t^2 + 8 \lambda^2)(\sqrt{t^2 + (4\lambda)^2} - t)^4 }{ 4096 \lambda^8} (U_\text{bdy} - U_\text{bulk}).  
	\end{eqnarray}
	With the full effective Hamiltonian given by $\hat H_\text{bdy} = \hat H_0 + \hat H_{U} $, we can take a continuum limit by sending the lattice constant to zero and obtain the helical Luttinger liquid
	\begin{eqnarray}
		\hat H_\text{bdy} = \frac{v}2 \left[ \Pi^2 + (\partial_x \varphi)^2 \right] + \frac{U_\text{eff}}{4\pi} \left[ - \Pi^2 + (\partial_x \varphi)^2 \right] =  \frac{\tilde v}2 \left[ K \Pi^2 + \frac1{K} (\partial_x \varphi)^2 \right] , 
	\end{eqnarray}
	where the renormalized velocity and Luttinger parameter are given, respectively, by, 
	\begin{eqnarray}
		\tilde v = v \sqrt{\left( 1 - \frac{U_\text{eff}}{2\pi v} \right) \left( 1 + \frac{U_\text{eff}}{2\pi v} \right)}, \quad K = \sqrt{\frac{ 1 - \frac{U_\text{eff}}{2\pi v}}{ 1 + \frac{U_\text{eff}}{2\pi v}}}.
	\end{eqnarray}
	Notice that the Rashba interaction leads to two-particle scattering between the two spin components, so after we include the Rashba interaction, the full boundary effective Hamiltonian is 
	\begin{eqnarray}
		\hat H_\text{bdy}  = \frac{\tilde v}2 \left[ K \Pi^2 + \frac1{K} (\partial_x \varphi)^2 \right] + g_2 \cos (2 \sqrt{4\pi} \varphi),
	\end{eqnarray}
	where $g_2 \propto V$. 
	Note that we used a different notation of the coupling constant $g_i$ compared to other bosonization conventions.

\end{document}